\documentclass[epj,nopacs,final]{svjour}
\usepackage{graphics}
\usepackage{latexsym}
\usepackage{subfigure}
\usepackage{epsfig,color,rotating,amsmath,delarray,array}
\usepackage{makeidx,pifont,float,amssymb}
\definecolor{gray01}{gray}{0.9}
\definecolor{gray02}{gray}{0.8}
\definecolor{gray03}{gray}{0.7}
\definecolor{gray04}{gray}{0.6}
\definecolor{gray05}{gray}{0.5}
\definecolor{gray06}{gray}{0.4}
\definecolor{gray07}{gray}{0.3}
\definecolor{gray08}{gray}{0.2}
\definecolor{gray09}{gray}{0.1}

\newcommand{\pir}{$\rm\gamma p\rightarrow p\pi^{0}$~}

\newcommand{\pirbf}{$\boldsymbol\gamma\textbf{p}\boldsymbol\rightarrow
\textbf{p}\boldsymbol\pi^{\textbf{0}}$~}

\begin{document}

\title{Photoproduction of \boldmath$\pi^0$\unboldmath $ $ Mesons
off Protons from the \boldmath$\rm\Delta(1232)$\unboldmath $ $ Region
to \boldmath$E_{\gamma}=3$\unboldmath\ GeV}
\titlerunning{Photoproduction of $\pi^0$ Mesons}
\authorrunning{The CB-ELSA Collaboration}
\author{
The CB-ELSA Collaboration \medskip \\
H.~van~Pee$\,^{1}$,
O.~Bartholomy$\,^1$,
V.~Crede$\,^{1,2}$,
A.V.~Anisovich$\,^{1,3}$,
G.~Anton$\,^4$,
R.~Bantes$\,^5$,
Yu.~Beloglazov$\,^3$,
R.~Bogend\"orfer$\,^4$,
R.~Castelijns$\,^{6,\rm a}$
A.~Ehmanns$\,^1$,
J.~Ernst$\,^1$,
I.~Fabry$\,^1$,
H.~Flemming$\,^{7,\rm b}$,
A.~F\"osel$\,^4$,
M.~Fuchs$\,^1$,
Ch.~Funke$\,^1$,
R.~Gothe$\,^{5,\rm c}$,
A.~Gridnev$\,^{3}$,
E.~Gutz$\,^1$,
St.~H\"offgen$\,^5$,
I.~Horn$\,^1$,
J.~H\"o\ss l$\,^4$,
J.~Junkersfeld$\,^1$,
H.~Kalinowsky$\,^1$,
F.~Klein$\,^5$,
E.~Klempt$\,^1$,
H.~Koch$\,^7$,
M.~Konrad$\,^4$,
B.~Kopf$\,^7$,
B.~Krusche$\,^8$,
J.~Langheinrich$\,^{4,\rm c}$,
H.~L\"ohner$\,^6$,
I.~Lopatin$\,^3$,
J.~Lotz$\,^1$,
H.~Matth\"ay$\,^7$,
D.~Menze$\,^5$,
J.~Messchendorp$\,^{9,\rm d}$,
V.A.~Nikonov$\,^{1,3}$,
D.~Novinski$\,^3$,
M.~Ostrick$\,^{5,\rm e}$,
A.~Radkov$\,^3$,
A.V.~Sarantsev$\,^{1,3}$,
S.~Schadmand$\,^{9,\rm a}$,
C.~Schmidt$\,^1$,
H.~Schmieden$\,^5$,
B.~Schoch$\,^5$,
G.~Suft$\,^4$,
V.~Sumachev$\,^3$,
T.~Szczepanek$\,^1$,
U.~Thoma$\,^{1,9}$,
D.~Walther$\,^5$,~and
Ch.~Weinheimer$\,^{1,\rm f}$ }

\institute{$^1\,$Helmholtz-Institut f\"ur Strahlen- und Kernphysik,
Universit\"at Bonn, Germany\\
$^2\,$Department of Physics, Florida State University, Tallahassee, FL,
USA\\
$^3\,$Petersburg Nuclear Physics Institute, Gatchina, Russia\\
$^4\,$Physikalisches Institut, Universit\"at Erlangen, Germany\\
$^5\,$Physikalisches Institut, Universit\"at Bonn, Germany\\
$^6\,$Kernfysisch Versneller Instituut,
Groningen, The Netherlands\\
$^7\,$Institut f\"ur Experimentalphysik I, Universit\"at Bochum,
Germany\\
$^8\,$Institut f\"ur Physik, Universit\"at Basel,
Switzerland\\
$^9\,$II. Physikalisches Institut, Universit\"at
Gie{\ss}en, Germany\\[1mm]
$^{\rm a}\,$Present address: Institut f\"ur Kernphysik,
Forschungszentrum J\"ulich, Germany\\
$^{\rm b}\,$Present address: GSI, Darmstadt, Germany\\
$^{\rm c}\,$Present address:
University of South Carolina, Columbia, SC, USA\\
$^{\rm d}\,$Present
address: Kernfysisch Versneller Instituut, Groningen, The Netherlands\\
$^{\rm e}\,$Present address: Institut f\"ur Kernphysik, Universit\"at
Mainz, Germany\\
$^{\rm f}\,$Present address: Institut f\"ur
Kernphysik, Universit\"at M\"unster, Germany\\
}
\date{Received: \today / Revised version:}

\abstract{Photoproduction of  $\pi^0$ mesons was studied with
  the Crystal-Barrel detector at ELSA for incident energies from
  300\,MeV to 3\,GeV. Differential cross sections d$\sigma$/d$\Omega$,
  d$\sigma$/d$t$, and the total cross section are presented. For
  $E_{\gamma}<3$\,GeV, the angular distributions agree well with the
  SAID parametrization. At photon energies above 1.5\,GeV, a strong
  forward peaking indicates $t$-channel exchange to be the dominant
  process. The rapid variations of the cross section with energy and
  angle indicate production of resonances. An
  interpretation of the data within the Bonn-Gatchina partial wave
  analysis is briefly discussed.  \vspace*{1mm}       
  }
  \mail{klempt@hiskp.uni-bonn.de}

\maketitle

%

\section{Introduction}
Due to their substructure, nucleons exhibit a rich spectrum of excited
states. A survey of the resonances observed so far can be found in
\cite{Yao:2006px}. In spite of considerable theoretical
achievements, attempts to model the nucleon spectrum with three
constituent quarks and their interactions still fail to reproduce
experimental findings in important details. In most quark-model based
calculations, more resonances are found than have been observed
experimentally \cite{Capstick:bm,Metsch}.  However, the quark model is
only an approximation and may overpredict the number of states
\cite{lichtenberg,Klempt:2004yz}. Alternatively, these {\it missing
resonances} could as well have escaped experimental observation due to a
weak coupling to N$\pi$ which makes them unobservable in elastic $\pi$N
scattering.

Resonances with small $\rm N\pi$ couplings are predicted to have sizable
photocouplings \cite{Capstick:bm}. Thus,  photoproduction of baryon
resonances provides an alternative tool to study nucleon states. New
facilities such as ELSA\,\footnote{\underline{EL}ectron
\underline{S}tretcher \underline{A}ccelerator} at Bonn,
Graal\,\footnote{\underline{Gr}enoble \underline{A}nneau
\underline{A}ccelerateur \underline{L}aser} at Grenoble, Jefferson Lab
(Virginia), MAMI C\,\footnote{\underline{MA}inz \underline{MI}crotron}
at Mainz, and SPring-8 (Hyogo) offer the opportunity to investigate
photoproduction for $E_{\gamma}>1$\,GeV and to study nucleon resonances
above the first and second resonance region.

Good angular coverage is needed to be able to extract the
resonant and non-resonant contributions in a partial wave
analysis. The Crystal-Barrel detector at ELSA is thus an ideal tool for
studying nucleon resonances.

Here, we present differential cross sections for the reaction
\begin{equation} \rm\gamma
p\rightarrow p\pi^0.
\label{EquationReaction}
\end{equation}
Nucleon resonances contributing to this reaction are not expected to
belong to the class of {\it missing} resonances but when searching for
these, identification of known resonances in photoproduction is an
important step.

The results on $\pi^0$ and $\eta$ photoproduction using the CB-ELSA
detector were communicated in two letters
\cite{Bartholomy:2004uz,Crede:2003ax}.  In this paper, we give full
account of the experiment and of data analysis of the reaction
$\rm\gamma p\to p\pi^0$. A publication covering all aspects related to
$\eta$ photoproduction is in preparation \cite{Bartholomy:2007}.

This paper is organized as follows: In section \ref{SectionOldData}, we
give a survey on the data already published before the CB-ELSA
experiment. The experiment itself is then described in section
\ref{SectionExperimentalSetup}. Section \ref{SectionData} provides a
detailed description of the data taken during the first data taking
periods and of the methods used in the event reconstruction. The
determination of the differential cross sections and the treatment of
systematic errors are discussed in section \ref{SectionCrossSections}.
Section \ref{SectionResults} contains a short description of the PWA
method used to extract the contributing
resonances from the data
\cite{Anisovich:2004zz,Anisovich:2005tf,Sarantsev:2005tg}. A summary is
given in section \ref{SectionSummary}.

\section{Previous Results on \boldmath$\pi^0$\unboldmath\
Photoproduction} \label{SectionOldData}

\begin{figure} \begin{center}
\epsfig{file=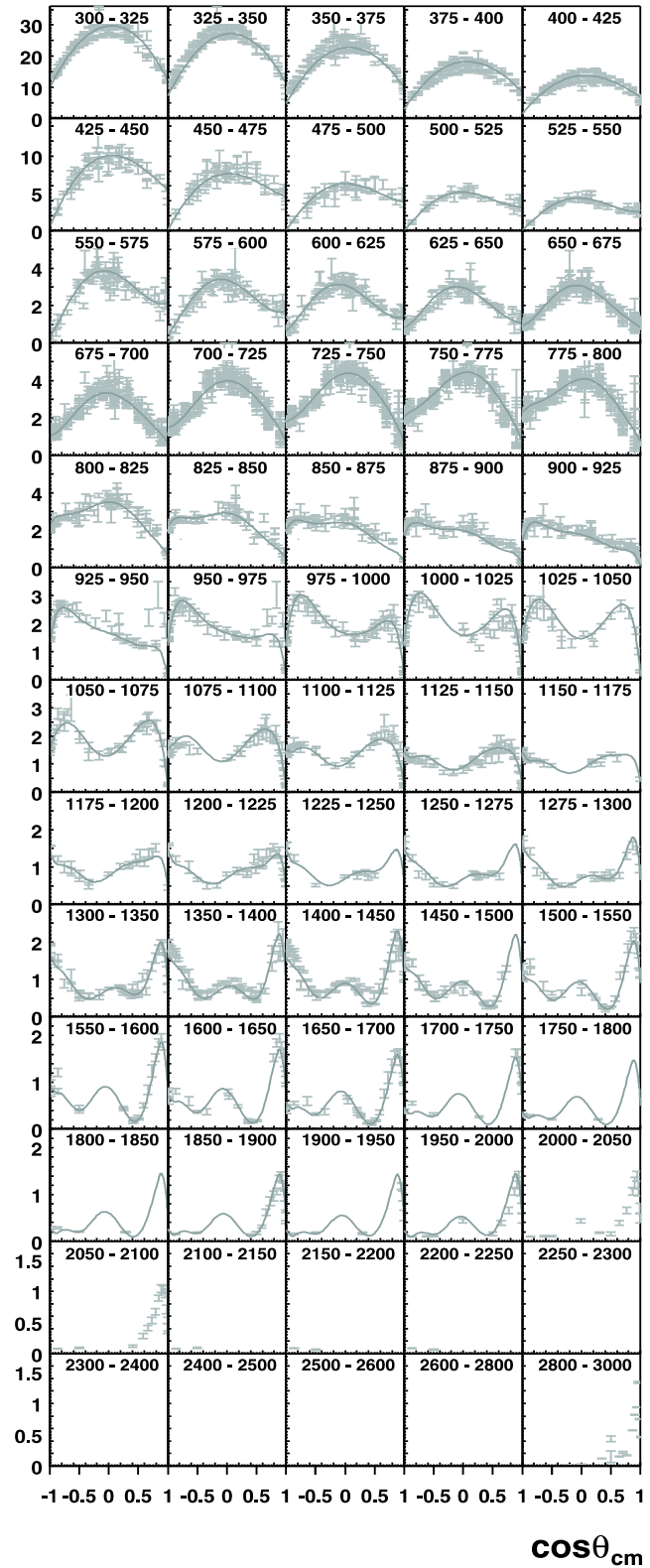,width=0.47\textwidth,height=0.92\textheight}
\caption{Differential cross sections $\rm d\sigma/d\Omega$ versus
  $\cos\Theta_{\rm cm}$ as measured by various experiments before
  CB-ELSA. The data stem from \cite{SAID}. The SAID model for the
  respective energy interval is shown as solid line. The photon energy
range in MeV is indicated in the subfigures.} \label{FigurePi0OldData}
\end{center} \end{figure}

First data on photoproduction of neutral pions date back to the 1960's.
The older experiments were limited in angular coverage and energies.
Fig.~\ref{FigurePi0OldData} shows the available differential cross
section data for the energy range from 0.3 to 3\,GeV. The data are
from the SAID\footnote{\underline{S}cattering \underline{A}nalysis
\underline{I}nteractive \underline{D}ial-In Program} compilation
\cite{SAID}.

In general, there is good coverage at lower energies. At beam energies
above 2\,GeV, there are only few data points measured by different
experiments at discrete angles.

The description of the data using the SAID model is also shown in
Fig.~\ref{FigurePi0OldData}. The SAID
model was fitted to these data points and thus, is only reliable up to
an incoming photon energy of about 2\,GeV.

In the following, a summary of the experiments performed and published
after 1979 is given.

Yoshioka {\it et al.} \cite{Yoshioka:1980vu} extracted differential
cross sections for photon-beam energies between 390 and 975\,MeV
in energy bins of 20 to 25\,MeV. They measured at 11
scattering angles covering a range in the center-of-mass system (cms)
of $15^{\circ}<\Theta_{\rm cm}<130^{\circ}$ corresponding to
$-0.64<\cos\Theta_{\rm cm}<0.97$.

At the Saskatchewan Accelerator Laboratory, differential cross sections
were measured for 11 energies within 25\,MeV above the
$\pi^0$-photoproduction threshold \cite{Bergstrom:1997jc}. The Igloo
spectrometer was used, which was especially designed for an excellent
$\pi^0$-detection efficiency. A full angular coverage from $0^{\circ}$
to $180^{\circ}$ was achieved.

Beck {\it et al.} measured differential cross sections at the electron
accelerator MAMI for five energy bins from threshold at 144\,MeV up to
photon-beam energies of 157\,MeV \cite{Beck:1990da} and for six further
bins between 270\,MeV and 420\,MeV \cite{Beck:1997ew}. Both experiments
used a linearly-polarized photon beam produced via coherent
bremsstrahlung. In \cite{Beck:1997ew}, the reaction was studied with
the DAPHNE\footnote{ \underline{D}\'etecteur \`a grande
\underline{A}cceptance pour la \underline{PH}ysique
\underline{N}ucl\'eaire \underline{E}xp\'erimentale} detector which
covers $\sim$\,94\,\% of the solid angle.

Differential and total cross sections were determined with TAPS\footnote{
\underline{T}wo \underline{A}rm \underline{P}hoton \underline{S}pectrometer}
at MAMI \cite{Fuchs:1996ja}. The TAPS collaboration measured at nine
different energies between threshold and 280\,MeV. The setup used five
blocks of crystals resulting in coverage of the full angular range in
$\Theta_{\rm cm}$, but only in partial coverage of the azimuthal angle.
The experiment covered about 50\% of the total solid angle. The
published data are restricted to energies below 152.5\,MeV. The
contribution of the $E_{0+}$ multipole was extracted and compared to
predictions from chiral perturbation theory and low-energy theorems.

Krusche {\it et al.}~\cite{Krusche:1999tv} extended the experiment to
cover the energy range up to 792\,MeV, which was the maximum possible
energy at MAMI. A comparison was made between $\pi^0$ photoproduction
off protons and off deuterons.

Schmidt {\it et al.}~\cite{Schmidt:2001vg} measured differential and total
cross sections for $\pi^0$ photoproduction for incoming photon energies
between threshold and 165\,MeV. The measured data points were then also
compared to predictions based on chiral perturbation theory and
low-energy theorems.

Ahrens {\it et al.}~\cite{Ahrens:2002gu} measured differential
cross sections for 12 energies in the photon-energy range between
550 and 790\,MeV using the DAPHNE detector at MAMI. The
availability of a polarized photon beam and a polarized target made it
possible to measure the helicity difference
$\sigma_{3/2}-\sigma_{1/2}$. The goal of the experiment was to test the
GDH\footnote{\underline{G}erasimov-\underline{D}rell-\underline{H}earne}
sum rule \cite{Gerasimov:1965et}.

At BNL\footnote{\underline{B}rookhaven \underline{N}ational
\underline{L}aboratory}, photoproduction cross sections for $\pi^0$ mesons
were measured using LEGS\footnote{\underline{L}aser \underline{E}lectron
\underline{G}amma \underline{S}ource}. Final-state particles were
detected in an array of six NaI crystals. The data cover the
beam-energy range from 213\,MeV to 333\,MeV. Unpolarized differential
cross sections as well as beam asymmetries were determined
\cite{Blanpied:2001ae}.

This compilation shows that (almost) all data taken after 1979 cover
only the lower photon-beam energy range up to 1\,GeV. Data above
1\,GeV stem from even older experiments; they are included in the SAID
database \cite{SAID}, too. The only recent publications comprising new
experimental data on $\pi^0$ photoproduction originate from the GRAAL
Collaboration~\cite{Bartalini:2005wx} and from this experiment
\cite{Bartholomy:2004uz}.

All available data (except~\cite{Bartholomy:2004uz,Bartalini:2005wx})
were reproduced well by the SAID model solution SM02~\cite{SAID}. Data
and fit results are shown in Fig.~\ref{FigurePi0OldData}. The
partial-wave analysis (PWA) also included $\pi$N-scattering data and
determined masses, widths, and photocouplings of baryon resonances. The
fit covered a photon-energy range up to 2\,GeV. Contributions of the
following resonances were extracted~\cite{Arndt:2002}:\\[2mm]
N(1440)P$_{11}$, N(1520)D$_{13}$, N(1535)S$_{11}$, N(1650)S$_{11}$,\\
N(1675)D$_{15}$, N(1680)F$_{15}$, $\Delta$(1232)P$_{33}$,
$\Delta$(1620)S$_{31}$,\\
$\Delta$(1700)D$_{33}$,
$\Delta$(1905)F$_{35}$, $\Delta$(1930) D$_{35}$,
and $\Delta$(1950)F$_{37}$. \\[2mm]
Recently, a new SAID model solution (SM05) has been released which, in
addition, takes into account the data presented in this paper, our data
on $\rm \gamma p\to p\eta$ \cite{Crede:2003ax} and the new GRAAL data
on \pir \cite{Bartalini:2005wx}.

The Mainz unitary isobar model MAID concentrates on the photon-beam
energy range below 1\,GeV~\cite{Drechsel:1999}. Extensions of MAID are
in preparation \cite{Tiator:Trieste}. Besides SAID and MAID there is a
large number of other approaches to describe photon- and pion-induced
production of mesons; a survey is given in the recent paper by
Matsuyama, Sato and Lee \cite{Matsuyama:2006rp}.

\section{Experimental Setup}
\label{SectionExperimentalSetup}

\subsection{Overview}

The CB-ELSA experiment was performed at the electron stretcher
accelerator ELSA. The maximum achievable electron energy is
3.5\,GeV.

The accelerator complex comprises three stages. A linear accelerator
(LINAC) pre-accelerates electrons emitted by a thermionic electron
gun to an energy of 20\,MeV. These are transferred to a booster
synchrotron, where they can be accelerated up to 1.6\,GeV forming a
pulsed beam ($f=50$\,Hz). The electron bunches are injected into the
stretcher ring over several cycles. The stretcher ring accumulates
these bunches and accelerates the electrons up to the required energy.
The electrons are made available for the CB-ELSA experiment via slow
resonance extraction, with a duty factor of nearly 90\%.

\begin{figure*}[pt]
\begin{center}
\epsfig{file=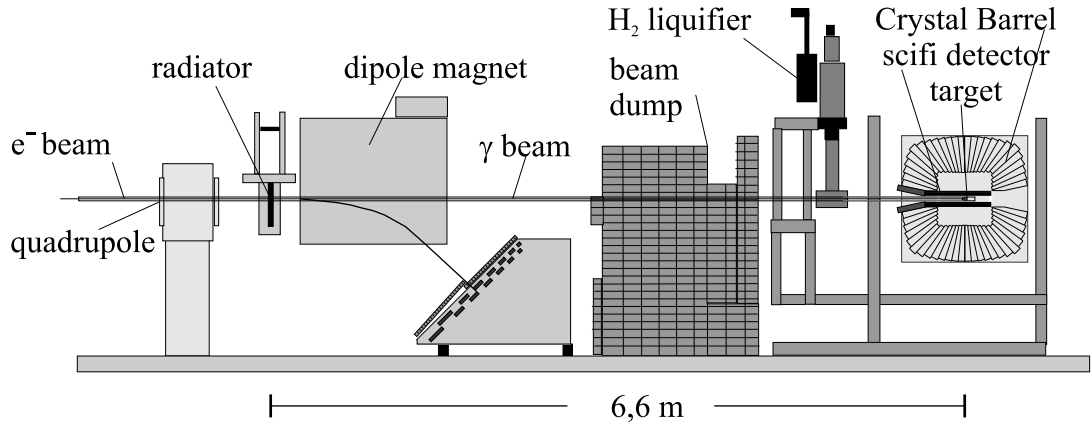,width=0.8\textwidth}
\caption{Experimental setup CB-ELSA, Bonn. The electron beam enters
  from the left side, hits the radiator, and produces bremsstrahlung.
  The photons are energy tagged and
  hit an LH$_2$ target in the center of the Crystal Barrel.
  Charged particles leaving the target are identified in
  the inner scintillating-fiber detector, photons are detected in the
  CsI(Tl) calorimeter. A photon counter for the flux determination is
  placed further downstream and is not shown in the figure.}
\label{FigureExperimentalSetup}
\end{center}
\end{figure*}

The experimental setup is shown in
Fig.~\ref{FigureExperimentalSetup}. A first short overview of the
different detector components is given below, which are then discussed
in more detail in the following subsections.

The electrons hit a radiator target, where they produce bremsstrahlung.
The energy of the photons, in the range between 25 \% and 95 \% of the
primary electron energy, was determined via the detection of the
corresponding scattered electrons in the tagging system.
The primary electron beam of unscattered electrons was stopped in a
beam dump situated upstream of the Crystal Barrel. The photon beam hit
the liquid hydrogen (LH$_2$) target (length: $l=52.84$\,mm, diameter:
$d=30$\,mm) placed in the center of the Crystal-Barrel detector.

The Crystal-Barrel detector forms the central component of
the experiment. It consists of 1380 CsI(Tl) crystals and has an
excellent photon-detection efficiency. The large solid-angle coverage
and the high granularity allow for the reconstruction of multi-photon
final states.  A more detailed description of the Crystal-Barrel
detector can be found in \cite{Aker:1992ny}.

Charged particles leaving the target cannot be unambiguously identified
by their energy depositions in the calorimeter. They were detected in a
three-layer scintillating fiber detector surrounding the
target~\cite{Suft:2005inner}. The first-level trigger of
the experiment exploited detection of protons; it was provided by the
tagging system in coincidence with the fiber detector. For the
photon-senstitive second-level trigger a \underline{FA}st
\underline{C}luster \underline{E}ncoder (FACE), based on cellular
logic, provided the number of clusters in the Crystal Barrel.
A segmented total-absorption oil $\check{\rm C}$erenkov counter (not
shown in Fig. \ref{FigureExperimentalSetup}) was placed further
downstream to determine the total photon flux traversing the target.

\subsection{Tagging System and \boldmath$\gamma$\unboldmath-Counter}
\label{SectionTagger}
The photon beam was produced by bremsstrahlung off an amorphous copper
foil of 3/1000 radiation length thickness. The total rate in the
tagging system during the beamtime was 1-3$\cdot 10^{6}$\,Hz. The
unscattered electron beam, deflected by 7$^{\circ}$, was
annihilated in a beam dump consisting of heterogeneous materials
including lead, boron-carbide, and polyethylene.

\begin{figure}[pb]
\begin{center}
\epsfig{file=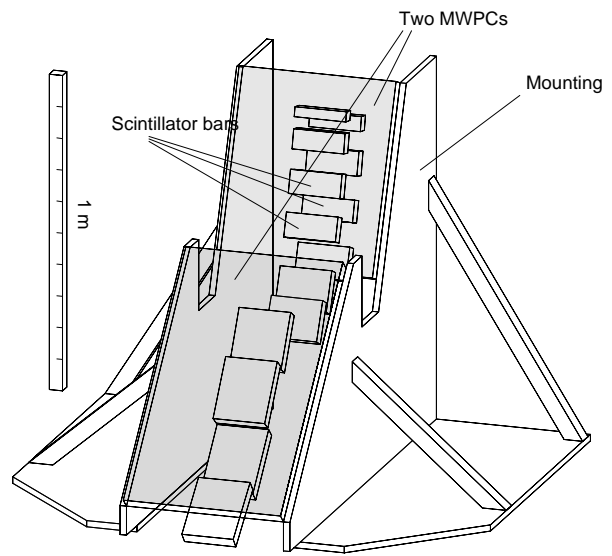,width=0.48\textwidth}
\caption{Layout of the tagging detector: The ladder of 14 scintillation
bars and two proportional wire chambers are shown.}
\label{FigureTagger} \end{center}
\end{figure}

The tagging system is shown in Fig.\,\ref{FigureTagger}. It serves to
analyse the electron momentum spectrum behind the radiator. According
to their energy loss, electrons are deflected in the dipole field of
the tagging spectrometer; the field strength (1 - 2\,T) was set
according to the chosen beam energy. Assuming single photon radiation,
the photon energy $E_{\gamma}$ is given by the beam energy $E_{\rm
e,~ELSA}$ and the energy of the detected electron $E_{\rm e}^{\prime}$,
\begin{equation}
E_{\gamma}\,=\,E_{\rm e,~ELSA}\,-\,E_{\rm e}'.
\label{EquationPhotonEnergy}
\end{equation}
Tagged photons are assigned to hadronic events in the
Crystal-Barrel setup by coincidences. Thus, the tagging system
consists of two distinct parts, a 4\,cm thick scintillator array to
provide fast timing information, and two
MWPCs\footnote{\underline{M}ulti-%
\underline{W}ire \underline{P}roportional \underline{C}hambers} with a
total of 352 channels (with four channels overlap). Their spatial
resolution translates into an energy interval per channel of 0.1
(0.5)\,MeV at the highest $E_{\gamma}$ and 10 (30)\,MeV at the lowest
$E_{\gamma}$ for the beam energies of 1.4 (3.2)\,GeV, respectively,
which were used for our measurements. The scintillator array was read
out via photomultipliers. A logical OR of the left-right coincidences
from all scintillators was required in the first-level trigger.

The tagging system was calibrated by direct injection of a very
low intensity $e^-$-beam of 600\,MeV or 800\,MeV, after removing the
radiator. Variation of the magnetic field of the tagging dipole
enabled a scan of several spatial positions over the MWPCs. For a given
wire, electron  momentum and magnetic
field strength are proportional. As long as saturation effects can be
ignored, the magnetic current is proportional to the field.

The calibration was checked by Monte-Carlo trace simulations of the
electron trajectories through the tagging magnet. Geometry of the
setup, dimensions of the electron beam, angular divergences,
multiple scattering and M$\o$ller scattering in the radiator foil and
the air were taken into account. From these simulations, the energy of
each of the MWPC wires was obtained by a polynomial fit. The
uncertainty of the simulation was estimated to be of the same order of
magnitude as the energy width of the respective wire. Deviations
between the 600\,MeV/800\,MeV calibration could be attributed to
effects of magnetic field saturation. Hence, the calibration of the
highest photon energy points relied on an extrapolation of the
simulated trajectories.

The hit-wire distribution of the MWPCs was measured with a {\it
minimum-bias} trigger at a fixed rate of 1\,Hz. This trigger required
only a hit in the tagging system and was thus independent of hadronic
cross sections. Accepted wire hits had to be isolated (one hit or
cluster of hits); the small background was estimated from the time
distribution of the associated tagger scintillator.

The absolute normalization was
treated as a free parameter which was determined by fitting the measured
angular distributions to the SAID cross sections. For the low-energy
data ($E_{\rm e,~ELSA}=1.4$\,GeV), the normalization constant was
determined for each energy bin, and an error of $\pm5\%$ was assigned
to it. The $E_{\rm e,~ELSA}=3.2$-GeV data cover a region for which no
SAID prediction is available. An energy-independent normalization
constant was determined from a comparison of our differential cross
sections with SAID for $0.8\leq E_{\gamma}\leq 1.7$\,GeV.  A systematic
error of $\pm 15$\% is estimated to account for possible variations of
the background across the tagger.

Downstream of the tagger and behind the Crystal-Barrel detector a total
absorption  oil $\check{\rm C}$erenkov photon counter was mounted. It
consists of three segments, each made of a hollow lead cylinder with 10
lead blades surrounded by mineral oil. The light produced by traversing
particles was detected by two photomultipliers per segment.

\subsection{Liquid H\boldmath$_2$\boldmath\ Target}
\begin{figure}[pt]
 \begin{center}
 \includegraphics[width=0.50\textwidth]{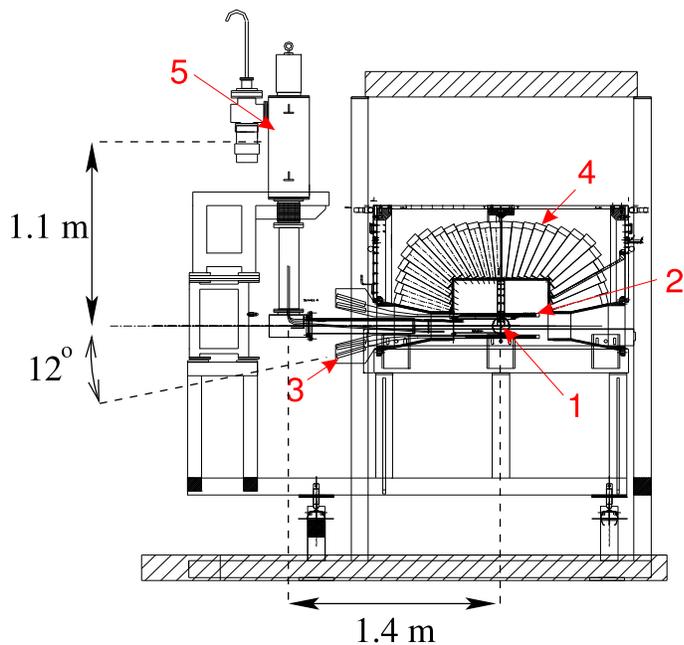}
 \caption{\label{cb_schnitt}
 The liquid hydrogen target with target cell (1),
 inner SciFi detector (2) and photomultiplier readout (3)
 are situated in the center of the Crystal-Barrel calorimeter (4).
 Due to the geometry, there is a large distance between  liquefier (5)
and target.
 }
 \end{center}
 \end{figure}
 Fig.~\ref{cb_schnitt} shows a side view of the assembly, the gas
liquefier, one half of the Crystal-Barrel calorimeter, target cell, and
inner scintillation fiber detector. The target cell cylinder is made of
125\,$\mu$m Kapton foil; entrance and exit window have a thickness of
80\,$\mu$m. The massive cold head of the H$_2$ refrigerator was mounted
2.5\,m outside of the detector to avoid obstruction of the detector
acceptance.

The {\it Air Products (model CSA-208-L)} refrigerator consists of a
two-stage cooler head operating in a closed helium circuit according
to the Gifford-McMahon principle.  In the first stage, H$_2$ is
liquefied. The second circuit consists of a large H$_2$ cold gas
reservoir and the target cell connected via two Kapton pipes of
75\,$\mu$m foil thickness serving as liquid  H$_2$ supply and
gaseous H$_2$ removal pipes.

\subsection{Crystal Barrel}
\begin{figure}[pt]
 \begin{center}
   \includegraphics[width=0.48\textwidth]{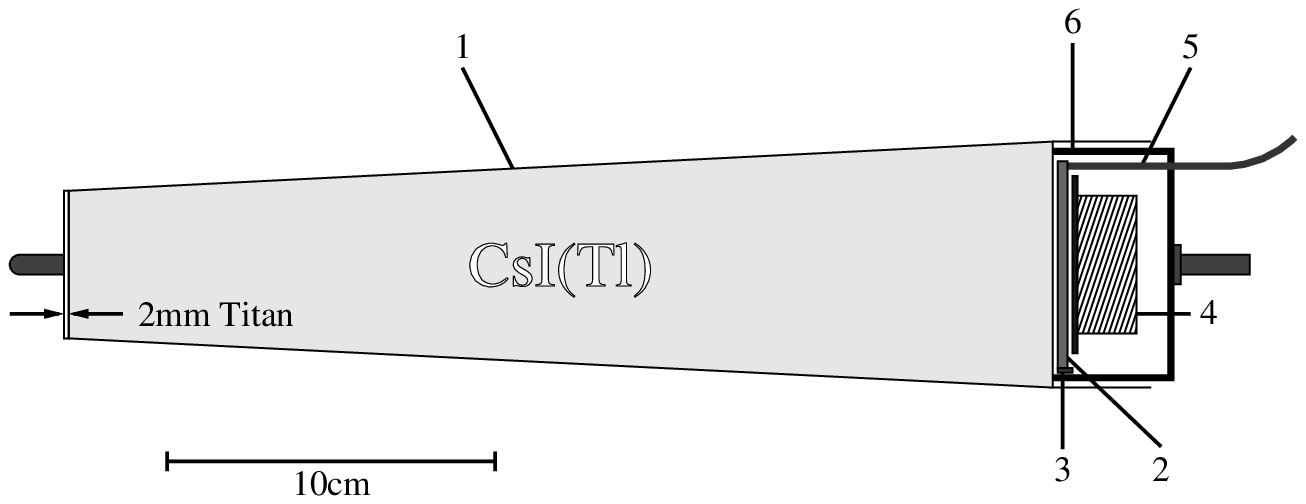}
 \end{center}
\caption{\label{cb_crystal}
 A crystal module: titanium case (1), wavelength shifter (2),
 photo-diode (3), preamplifier (4), optic fiber (5), case cover
 (6).
}
 \begin{center}
  \includegraphics[width=0.48\textwidth]{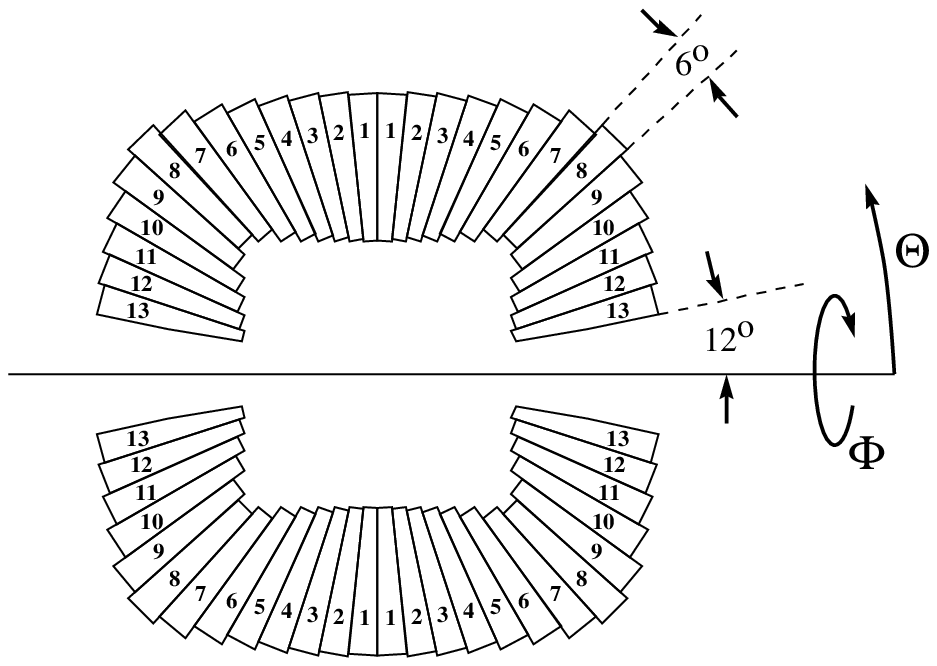}
 \caption{Mounting scheme of the Crystal-Barrel
 calorimeter. The numbers indicate the different crystal shapes.}
 \label{cb_calorimeter}
 \end{center}
 \end{figure}

The Crystal-Barrel calorimeter is designed
to provide high-efficiency photon detection with good energy and
spatial resolution over an energy range from 20\,MeV to 2\,GeV. All
crystal modules are arranged in a vertex-pointing geometry forming the
shape of a barrel. The Crystal Barrel consists of 1380 CsI(Tl) crystals
with a length of 30\,cm corresponding to 16 radiation lengths and
covering almost the complete solid angle. Each of the crystals is
wrapped with a 0.1\,mm titanium foil for mechanical stability and
protection and a 2\,mm support foil at the end (see Fig
\ref{cb_crystal}).

Fig.~\ref{cb_calorimeter} shows a schematic diagram of the crystal
arrangement.
The special shape of the calorimeter requires 13~different types of
crystals. They are arranged in 26~rings. Each ring has either
60 or 30~crystals, i.e. a single crystal covers $6^\circ$
(rings 1-10) or $12^\circ$ (rings 11-13) in azimuthal angle $\Phi$.
In polar angle, a range from 12$^\circ$ to~168$^\circ$ is covered
corresponding to a solid angle of~97.8\% of 4$\pi$.

The size of the crystal segments was adjusted to the Moli\`ere
radius of CsI crystals, $r_{\rm M} = 3.5$\,cm. The 12$^\circ$~opening on
either side of the barrel is necessary for technical reasons. The
segmentation limits the spatial resolution to 20\,mrad ($\approx
1.1^\circ$) in~$\Phi$ and~$\Theta$. However, it allows the separation
of two photons stemming from the decay of a $\pi^0$~meson with a
maximum momentum of~1\,GeV/$c$ corresponding to a minimum opening angle
of~16.6$^\circ$. The energy resolution of the calorimeter is
empirically described by
\begin{equation}
\frac{\sigma_E}{E}\;=\;\frac{2.5\%}{\sqrt[4]{E\,[{\rm GeV}]}}\,.
\end{equation}
For the readout of the scintillation light, silicon photodiodes are
used. They are placed on a wavelength shifter (WLS) mounted on the
rear end of the crystals and fulfilled two purposes.  The WLS collect
the light from the crystals over their full cross sectional area and
transform the wavelength of the detected light to the sensitive range
of the photodiode. CsI crystals emit light at wavelengths between 450
and 610\,nm, with a maximum at 550\,nm. The absorption profile of the
WLS exhibits a maximum between 520 and 590\,nm matching the range of
emission of CsI.  The light emitted by the WLS has a wavelength of 600
to 700\,nm. For this wavelength, the WLS is practically transparent and
the photodiode is most sensitive.

Preamplifiers are attached to the photodiodes on the crystal module to
reduce the background noise. The overall gain stability of the CsI
crystals is monitored by a light pulser system. For this purpose each
WLS is connected via an optical fiber to a common Xenon flashlamp
operating at a repetition rate of~5\,Hz. The light is read out by the
normal calorimeter electronics. Thus, crystal modules not working
properly can easily be identified. In addition, the light pulser
system is used to calibrate the high and the low range of the
12bit-dual range ADCs.

The Crystal Barrel was already used successfully for seven years at
CERN\footnote{\underline{C}onseil \underline{E}urop\'een pour la
\underline{R}echerche \underline{N}ucl\'eaire}
(LEAR\footnote{\underline{L}ow-\underline{E}nergy \underline{A}ntiproton
\underline{R}ing}) before it was brought to Bonn. Changes of the photon
yield in the crystals due to radiation damage were not observed. A
detailed description of the CERN detector can be found in
\cite{Aker:1992ny}.

The calibration of the Crystal Barrel was carried out after data taking.
For each crystal, the $\pi^0$ peak in the two-photon invariant mass was
normalized iteratively to the nominal mass of $m_{\pi^0}=134.98$\,MeV by
taking the invariant masses of each pair of two photons. The mass
resolution is $\sigma=8$\,MeV at $m_{\pi^0}$ and $\sigma=15$\,MeV at
$m_{\eta}$ in their $2\gamma$ decays, and $\sigma=20$\,MeV at
$m_{\omega}$ (in its $\pi^0\gamma$ decay).

\subsection{Inner Scintillating-Fiber Detector}
The three-layer scintillating fiber detector identifies charged
particles leaving the target and determines their intersection point
with the detector \cite{Suft:2005inner}. The 40\,cm
long detector is mounted on a 1.8\,mm~thick aluminum support
structure. The fibers are 2\,mm~in diameter; they are positioned
at mean radii of~5.81\,cm, 6.17\,cm and~6.45\,cm, respectively.
The innermost layer corresponds to a solid angle of~92.6\%~of~4$\pi$,
thus covering almost the full solid angle of the barrel detector. The
fibers of the outer layer were installed parallel to the z-axis,
the fibers of the inner layer are bent clockwise, the remaining fibers
are bent anti-clockwise, forming respectively, an angle of $-24.5^\circ$
or 25.7$^\circ$ with the z-axis. Between each pair of layers, carbon
cylinders hold the fibers in place. In total, the detector
consists of 513 scintillating fibers read out via
16-channel-photomultipliers.  Each scintillating fiber is
coupled individually to an optical fiber guiding the signal through the
backward opening of the barrel to the photomultipliers.
The 3 layers had efficiencies of $94.8\pm 0.9$\% (inner), $92.9\pm
0.9$\% (middle), and $88.1\pm 0.8$\% (outer), respectively. These
values include the geometrical acceptance. The probability of two out
of three layers having fired was $98.4\pm 0.2$\% \cite{Suft:2005inner}.

\subsection{Trigger and Data Acquisition System}
The first-level trigger demanded a coincidence of a hit in one of the
tagger scintillators (related to an incoming photon) and of hits
in two or three different layers of the inner scintillating-fiber
detector (interpreted as detection of a proton in the final state). The
tagger rate was of the order of 1-3$\cdot 10^6$\,Hz; the coincidence
with scintillating-fiber hits reduced this rate to 2000\,Hz. The
second level trigger defined the number of contiguous clusters of hit
crystals. It used a fast cluster encoder (FACE) based on cellular
logic. The decision time depended strongly on the complexity of the hit
distribution in the Crystal Barrel and was typically 4\,$\mu$s. In
case of rejection of the event, a fast reset was generated, which
cleared the readout electronics in 5\,$\mu$s. Otherwise the readout
of the full event was initiated; typical readout times were 5-10\,ms.
Along with an incoming trigger rate of 2000 Hz from the inner
detector, this led to a dead time of 70\% when two clusters were
required by FACE. For three clusters in FACE, the dead time was
negligible.

For the data presented here, events were only recorded if they had more
than two clusters determined by FACE from the pattern of hit crystals.
For part of the data the minimum number of clusters was set to three.
The average overall data taking rate of the DAQ was 100\,Hz or more,
depending on whether empty channels were suppressed or not.

\section{Data Analysis}
\label{SectionData}
The data presented here were taken from December 2000 until July 2001
in two run periods with different primary electron energies of 1.4 and
3.2\,GeV. In the following, we refer to these data sets according to
their incident electron energies. Cross sections were determined
separately for the two different settings providing a total range of
photon energies from 0.3 to 3\,GeV. This corresponds to a range of
$\gamma$p or p$\pi^0$ invariant masses from 1.2 to 2.6\,GeV/$c^2$. The
$\pi^0$ is reconstructed from its $\pi^0\to\gamma\gamma$ decay.

\subsection{Photon Reconstruction in the Crystal Barrel}

An electromagnetic shower will, in general, extend over several crystals.
Such an area of contiguous calorimeter modules is called a ``cluster''.
In order to reduce the contributions of noise to the cluster energy,
only crystals with an energy deposit above a threshold of~1\,MeV are
considered in the cluster finding algorithm. Within a cluster, a search
was made for local energy maxima with an energy deposition above a
minimum value of 20\,MeV~($E_{\rm CLS}$). A crystal containing a local
maximum of energy is referred to as a {\it central crystal}. Any local
maximum is interpreted as evidence that a (charged or neutral) particle
has hit the Crystal Barrel. Its energy deposit in the Barrel is called
PED (particle energy deposit). If a cluster contains only one local
maximum, the sum over all crystal energies in the cluster was then
assigned to the energy of the PED ($E_{\rm PED}$). The center of
gravity $\vec{x}=(\Theta,\Phi)$ of the energy distribution defines the
spatial coordinates of the impact point on the Crystal-Barrel surface,
and the momentum of a photon:
\begin{equation} \vec{x} = \frac{\sum_j \left(P +  \ln\left(
\frac{E_{j}}{E_{\text{PED}}}\right)\right) \vec{x}_{j}} {\sum_j \left(P
+  \ln \frac{E_{j}}{E_{\text{PED}}}\right)}
\quad
\end{equation}
\begin{equation}
\text{ for }\quad\frac{E_j}{E_{\text{PED}}}\ge e^{-P}\,,
\end{equation}
where $j$ runs over all crystals in the cluster and the cut-off
parameter is $P = 4.25$. A logarithmic weight has a stronger impact
from small (and more distant) energy contributions leading to a better
reconstruction accuracy of the photon direction. The weighting
procedure was optimized performing extensive Monte-Carlo simulations.
The PED is identified with a photon if it cannot be associated in space
with the projection of a charged track (see below) extrapolated to the
surface of the Crystal Barrel. Otherwise, it is identified as charged
particle and here, by default, as proton.

If more than one local energy
maximum is found inside a cluster, the energy of PED $i$ is
determined from the energy deposit of the central crystals and the
deposits of the up to eight neighbors, summed up to form $E_9^{i}$. The
total energy of the cluster is then shared between the PEDs
$j=1,2,\cdots$  according to the relative magnitude of their
$E_9$-sums:
\begin{equation} E^i_{\mbox{PED}}\;=\;\frac{E_9^i}{\sum_j
E_9^j}\;E_{\mbox{cluster}} \quad {\mbox{and}} \quad
E_9\;=\;\sum_{i=k}^9\;E_k\;,
\end{equation} where crystal energies in
overlapping regions are split according to the relative energy deposits
of the local maxima.

Monte Carlo simulations showed that the reconstructed photon energies
differ from the initial values due to energy loss in insensitive
parts of the detector. A PED-energy- and $\Theta$-dependent function
derived from MC~simulations was applied to the reconstructed energies.

Statistical shower fluctuations can generate additional maxima which
are called {\it split-offs\/} and may be misidentified as photons. A
single photon can create several split-offs, so that some good events
are not considered in the analysis due to a wrong PED~multiplicity in
the barrel. About~5\%~of all photon showers create such split-offs.
Split-offs can also be generated by charged particles, e.g. by nuclear
reactions. The Monte Carlo simulation reproduces split-offs rather
precisely. The error originating from split-offs is estimated to be
less than 2\% \cite{Amsler:1993kg}.

\subsection{Identification of Charged Particles}
The event reconstruction in the Crystal Barrel provides direction and
energy of a photon ($\Phi,~\Theta$,~and~$E~+$ with errors). High-energy
charged particles like protons and pions can traverse the whole length
of a crystal module without depositing all their kinetic energy in the
CsI crystals. Pions originating in the target center and traversing the
Crystal Barrel show a minimum ionizing peak at about 170\,MeV. Their
kinetic energy cannot be deduced from their energy deposit in the
Crystal Barrel. The energy deposit of protons is larger and reaches a
maximum for those stopping at the rear end of the Crystals. Protons
stopping earlier and penetrating protons deposit less energy. A given
PED energy of a proton corresponds therefore to two allowed values for
the proton kinetic energy. In this analysis, the proton PED energy was
not used.

Identification of charged PEDs was important to perform the
analysis even though proton PEDs were ignored after identification.
Charged particles produce usually single-crystal
clusters, and the reconstructed angles $\Theta$ and $\Phi$ agree in
most cases with a crystal center, leading to regular spikes in the
proton angular distribution.  Therefore, protons were treated as
missing particles in a kinematic fit.

\begin{figure}[pt]
\begin{center}
\begin{tabular}{ccc}
\includegraphics[width=15mm]{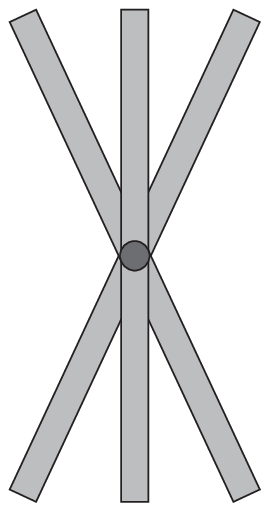}&
\hspace{3mm}\includegraphics[width=17mm]{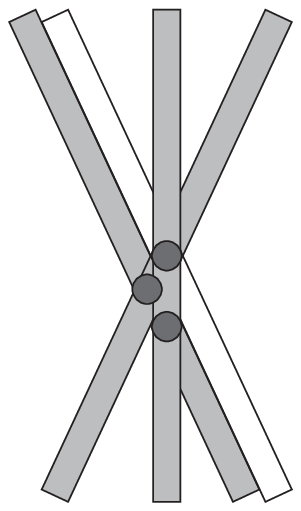}&
\hspace{3mm}\includegraphics[width=25mm]{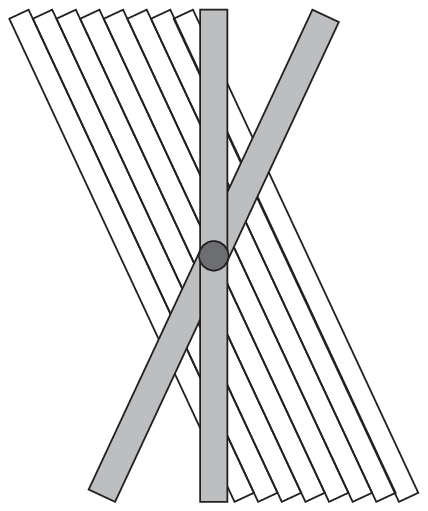}
\end{tabular}
\end{center}
\caption{Charged particles crossing the inner fiber detector produce
different hit patterns. A hit in three layers may result in one
unique intersection point or in three intersection points (middle).
One inefficiency still yields a defined intersection point (right).
} \label{kreuzung}
\end{figure}

A charged particle traversing the fiber detector can fire one or two
adjacent fibers in each layer forming a {\it British-flag}-like
pattern which defines the impact point (see Fig. \ref{kreuzung}, left).
A single charged particle may also fake three intersection points
(see Fig. \ref{kreuzung}, middle). In order not to loose these
events, up to three intersection points were accepted in the
reconstruction. One inefficiency (one broken or missing fiber) leading
to a pattern shown in Fig. \ref{kreuzung}, right, was accepted by the
reconstruction routines.

The accuracy of the reconstruction of the impact point was studied
using simulations and was determined to $\pm0.5$ mm in the x- and y-
coordinates (representing the resolution in $\Phi$). The error in the z
coordinate was determined to be 1.6\,mm.
Trajectories were formed from the center of the hydrogen target to the
impact point in the fiber detector as well as to all PEDs in the
barrel. For all resulting combinations, the angle between the
trajectories was calculated. A PED in the Crystal Barrel was
identified as a charged particle if the angle between a trajectory
starting in the target center and hitting the fiber impact point and
the trajectory from target center to the Crystal-Barrel PED was
smaller than $\approx 20^\circ (\,=0.35~{\rm rad})$. This procedure is
called matching.

In some cases, there was more than one hit per layer. If the hits
belonged to adjacent fibers, the clusters were assigned to the central
fiber (two wires give an effective fiber number $(2n+1)/2$); otherwise,
more than one impact point was reconstructed. Each impact point was
tested if it can be matched to a PED in the Crystal Barrel. Matched
PEDs are identified as protons. Only one proton is allowed in the event
reconstruction. Events with two matched PEDs, possibly due to $\rm
p\pi^+\pi^-\pi^0$ with one undetected charged particle, were rejected.

Protons going backwards in the center-of-mass system have rather low
momenta. The minimum momentum for a proton to traverse the target cell,
wrappings, inner detector, and Crystal-Barrel support structure and to
deposit 20\,MeV in a crystal is 420\,MeV/$c$. To reconstruct
events with protons having smaller momenta, events with two
PEDs in the Crystal Barrel were taken into account when they had
a hit in the inner detector not matching one of the two PEDs. It was
then assumed that the two PEDs were photons and that the proton got
stuck in the inner detector or support structure between
scintillation fiber detector and the barrel. Thus, proton
momenta down to 260\,MeV/$c$ were also accessible in the analysis.

\subsection{Event Reconstruction}

In the first step of data reduction, both photons from a $\pi^0$ decay
were required to be detected in the Crystal Barrel, irrespective of the
detection of the proton. Thus, events with 2 or 3 PEDs in the Crystal
Barrel were selected. If a proton candidate was found in the fiber
detector, and the fiber hit could be matched to a barrel hit, the
corresponding PED was identified as proton and excluded from
further analysis.  The proton was thus treated as missing particle.
Unmatched PEDs were considered to be photons. All events with two
photons were kept for further analysis. In this way, consistency
between the two data sets (with and without observed proton) was
guaranteed.

In total, $\sim$45 million events were assigned to these two event
classes and subsequently subjected to a kinematic fit as
described in the following.

Kinematic fits minimize the deviation of measured quantities from
constraints like energy and momentum conservation. The measured
values $y_i$ are varied (by $\delta y_i$) within the estimated
errors $\sigma_i$ until the constraints are fulfilled exactly. Further
constraints are given by the known masses of particles like $\pi^0$ or
$\eta$ which are reconstructed from the 4-momenta of photons. Kinematic
fitting improves the accuracy by returning corrected
quantities, which fulfill the constraints exactly. A $\chi^2$ value can
be calculated from the minimization given by
\begin{equation}
\chi^2\;=\;\sum_i\:\left(\frac{\delta y_i}{\sigma_i}\right)^2\;.
\end{equation}
In our experiment, the quantities $y_i$ are the measured
values $(\Phi, \Theta,\sqrt{E})$.
Since a fraction of the energy can be lost (but
never created) in some material, the distribution of $E$ is asymmetric;
therefore $\sqrt E$ is chosen as variable which exhibits a
more Gaussian distribution than $E$.

If the $y_i$~measurement errors are
correlated, then this becomes a matrix equation. Each constraint
equation is linearized and added, via the Lagrange multiplier
technique, to the $\chi^2$~equation:
\begin{equation}
\chi^2\;=\;(\delta\vec{y}\,)^T {\bf V}^{-1}(\delta\vec{y}\,)\:
+\:2\vec a\,^T\vec{f}(\vec{y}+\delta\vec{y}\,)\;,
\end{equation}
where~${\bf V}$~is the covariance matrix for the measurements
$\vec{y}=(y_i)$; \boldmath$\vec{a}$\unboldmath\ is the vector of
Lagrange multipliers for the constraints $\vec{f}$, and $\delta\vec{y}$
denote the required variations. When the constraining equations are
satisfied and the experimental error matrix correctly determined, then
the difference between true values $y_i$ and the measurements
$y_i+\delta y_i$ should be of the same magnitude as the
errors. If the
data are distributed according to a Gaussian function, a probability
distribution for the $\chi^2$~with $n$~degrees of freedom can be
defined by \begin{equation}
P(\chi^2;n)\;=\;\frac{2^{-n/2}}{\Gamma(n/2)}\:
\chi^{n-2}\:{\mbox{e}}^{-\chi^2/2}\;,
\end{equation}
which is called the {\it $\chi^2$~probability}. To make judgments and
decisions about the quality of the fit, the relevant quantity
is the integral
\begin{equation}
CL(\chi^2)\;=\;\int_{\chi^2}^\infty P(\chi^2;n)\:d\chi^2\;,
\end{equation}
which is called the {\it confidence level}.
The $\chi^2$~calculation depends directly on the errors of the measured
values. In this case, the errors are the measurement errors of the
calorimeter hits: the polar angle~$\Theta$, the azimuthal angle~$\Phi$
and the energy~$E$ of a calorimeter cluster. If the errors and, hence,
the covariance matrix are correctly determined, the confidence level
distribution of the fits should be flat. A sharp rise near zero of the
confidence level distribution can indicate contributions from
background events. The kinematical fit is thus an ideal method to
quantify the quality of different final-state hypotheses for an event.
\begin{figure}[pb]
\begin{center}
\epsfig{file=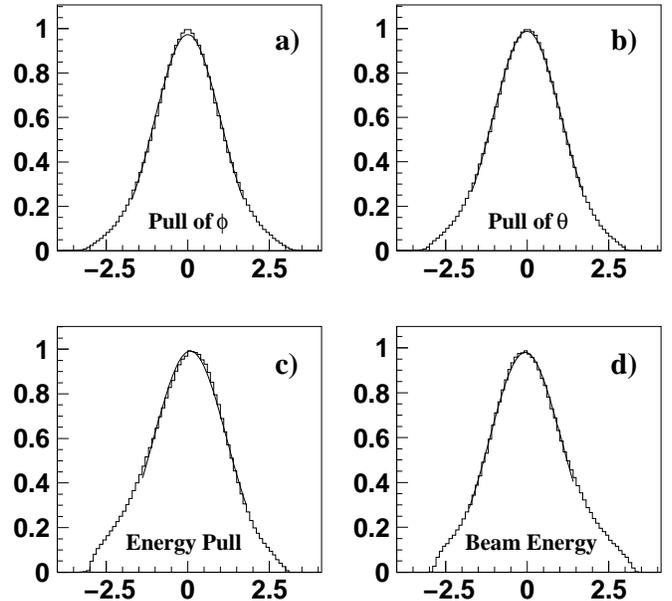,width=0.48\textwidth,height=0.44\textwidth}
\vspace*{2mm}
\caption{Deviations between the measured values $\Phi$ (a), $\Theta$
(b),  $\sqrt{E}$ (c) and $E_{\gamma}$ (d) as the results from a
kinematic fit in units of the respective measurement errors (1.4\,GeV
data). After calibration, the mean values should be 0 and the variance
1. The experimental distributions are compatible with these numbers. }
\label{FigureGauss} \end{center} \end{figure}

A {\it pull} is a measure of the displacement of the measured values to
the fitted values normalized to the corresponding errors. The measured
values $y_i$ with errors $\sigma_i$ are corrected by shifts $\delta y_i$
leading to new values $y_i+\delta y_i$ which fulfill exactly the
constraints and have smaller errors ${\sigma'}_i$. The quantity
\begin{equation}
{\rm pull_i} = \frac{\delta y_i}{\sqrt{\sigma_i ^2-{\sigma'}_i ^2 }}
\end{equation}
is called pull; it should be Gaussian with unit width and centered
around zero. Fig.~\ref{FigureGauss} shows the pulls for the Crystal
Barrel quantities ($\Phi$, $\Theta$, $\sqrt{E}$) and the photon beam
energy $E_{\gamma}$.  Mean values are all compatible
with zero, the variances are close to (but slightly above) 1.

 A systematic offset in a
measured quantity will move the center of its pull distribution away
from zero. Using this information in addition to kinematic fits
treating the z vertex as free parameter, an (unwanted) target
displacement of 0.65\,cm from the central position towards the tagging
system was detected, in perfect agreement with a later position
measurement.

\begin{figure}[pb]
\begin{tabular}{cc}
\hspace{-3mm}\epsfig{file=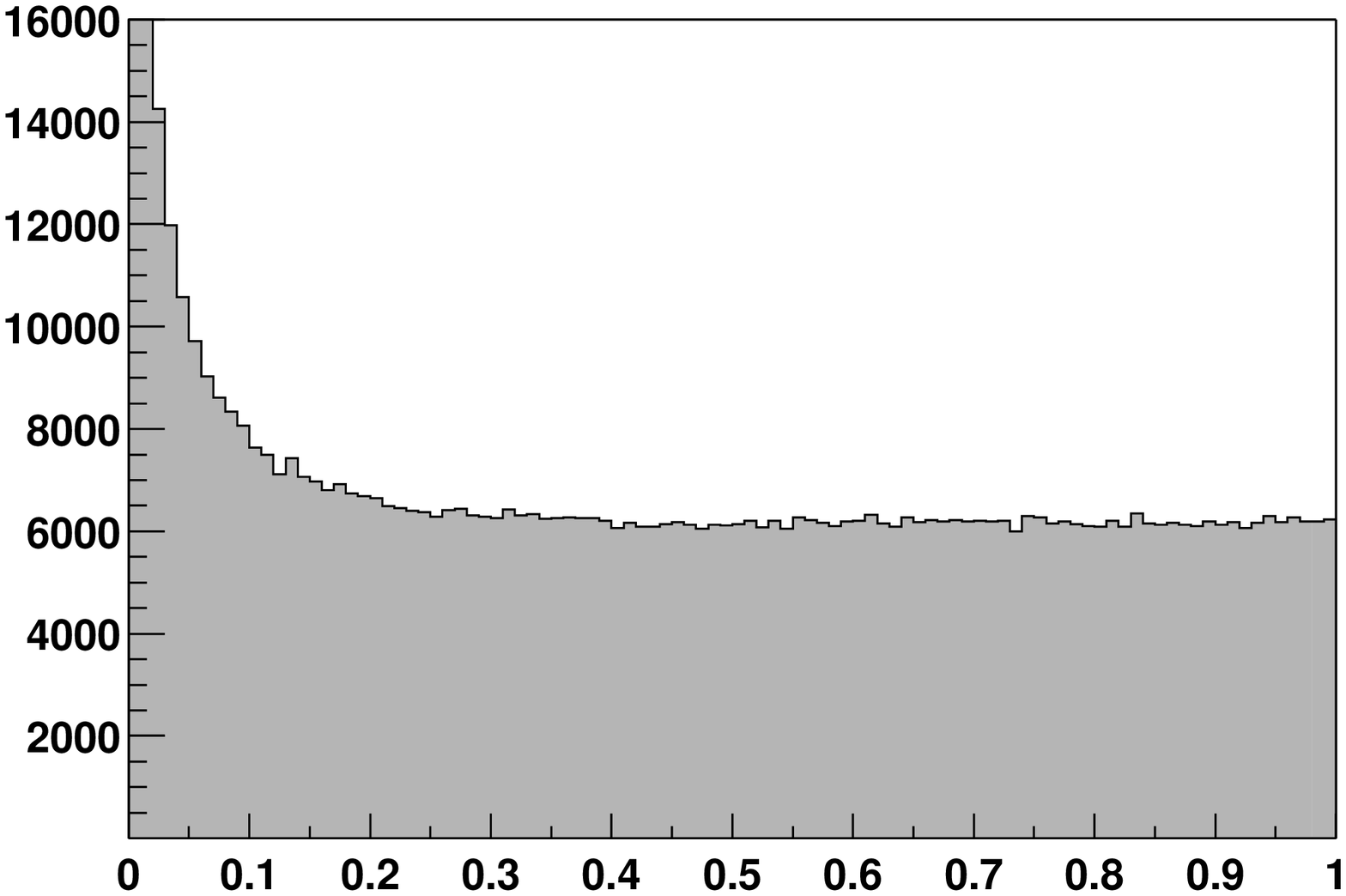,width=0.24\textwidth,height=0.24\textwidth}&
\hspace{-2mm}\epsfig{file=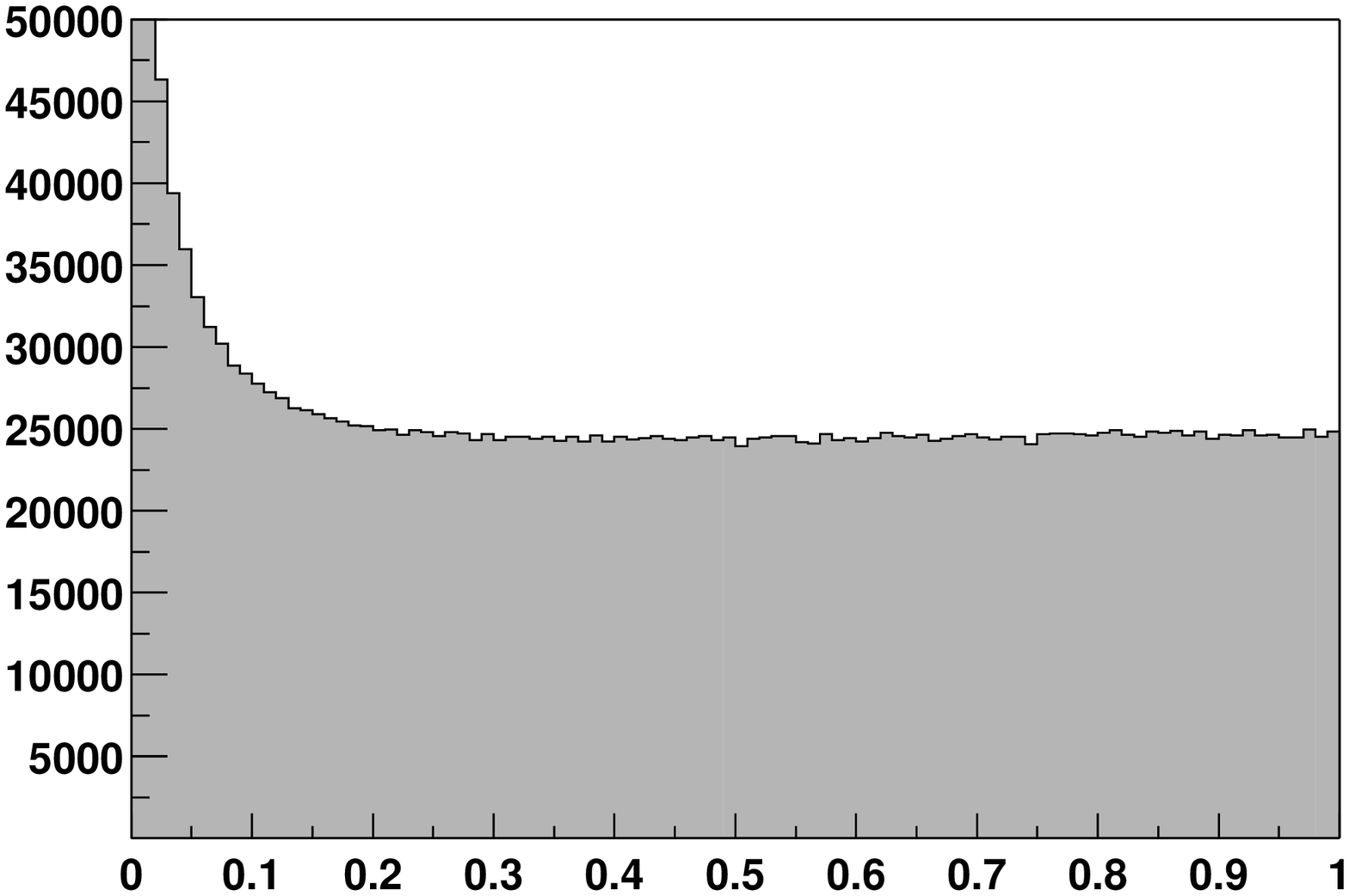,width=0.24\textwidth,height=0.24\textwidth}
\end{tabular}
\caption{\label{fig:cl}Confidence level distributions (number of
events as a function of CL) for the one-constraint kinematic fit to the
hypothesis $\rm \gamma p \rightarrow p 2\gamma$ imposing energy and
momentum conservation and treating the proton as unseen particle for
1.4-GeV data (left) and the corresponding Monte Carlo simulations
(right). } \end{figure}

In a preselection, the compatibility of events with the hypothesis
\begin{equation}
\rm \gamma p \rightarrow p 2\gamma
\end{equation}
was tested in a one-constraint (1C) kinematic fit imposing energy and
momentum conservation but leaving the proton 3-momentum as adjustable
quantity. The confidence level distribution of the fit is shown in
Fig.~\ref{fig:cl}. Above 20\%, the confidence level distribution is
flat in both data and Monte Carlo events. At
small confidence level CL, the distribution increases when CL
approaches zero in both distributions. In the data, this could indicate
the influence of background events or of too small errors for a
subclass of events. In the Monte Carlo simulation shown in
Fig. \ref{fig:cl}, only p$\pi^0$ events were created and no background
events. Hence the events with small confidence level must be true
p$\pi^0$ events. Indeed, inspection of $\gamma\gamma$ invariant mass
distribution shows that these events are $\rm p\pi^0$ events with
larger errors.  The assumption made in the
kinematic fit that the errors do not depend on the kinematics of the
event is obviously wrong. A cut at e.g. 10\% confidence level entails
therefore the risk that events with larger errors, preferentially
events with $\pi^0$ in forward direction, have only a small detection
efficiency. This would lead to a loss of statistics and, in case
of differences between data and simulations, to wrong results.

Here, a cut on the confidence level at $10^{-4}$ was applied. This cut
rejects all events where the kinematic fit fails (CL=0) and very
badly measured events. It was checked that the rejected events have a
$\gamma\gamma$ mass distribution with very few $\pi^0$ only.

\begin{figure}[pt]
\begin{center}
\subfigure[1.4\,GeV~data]{\epsfig{file=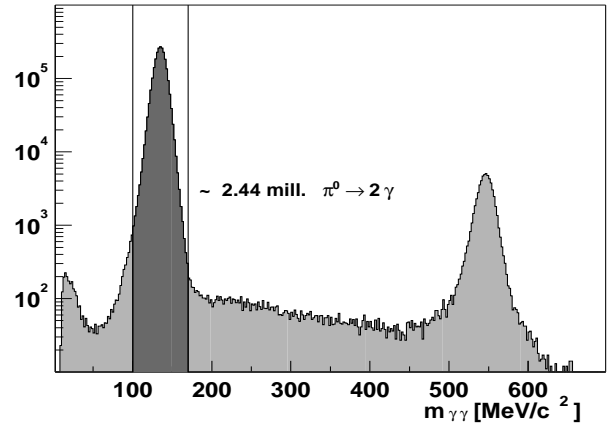,
width=0.44\textwidth,height=0.24\textheight}}
\vspace*{-3mm}\\
\subfigure[3.2\,GeV~data]{\epsfig{file=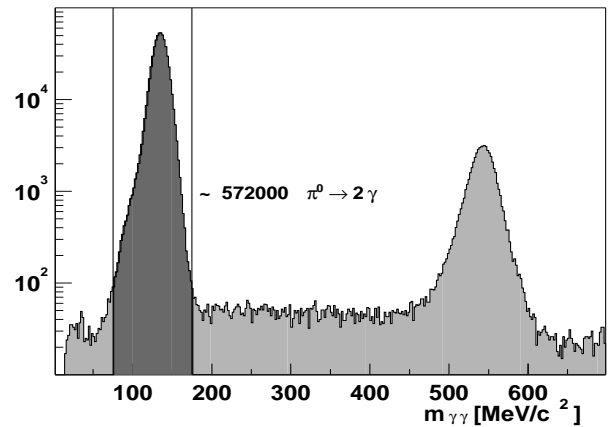,
width=0.44\textwidth,height=0.24\textheight}}
\vspace*{-2mm}\\
\caption{Two-photon invariant mass spectra using events with two
  properly reconstructed photons. Note the logarithmic scale. A
kinematic fit to the
  $\rm \gamma p\rightarrow p\gamma\gamma$ hypothesis was performed and
  a confidence level cut of $10^{-4}$ applied.  The dark grey area,
  defined by an additional mass cut, corresponds to accepted events.
   The low-mass asymmetry of the $\pi^0$ peak is due to energy
  overflow   in the ADCs for high-energetic photons. It is well
  reproduced by Monte Carlo simulations.}
\label{FigureGGMass}
\end{center}
\end{figure}

Fig.~\ref{FigureGGMass} shows $\gamma\gamma$ invariant mass
spectra after the confidence level cut: (a) for
the low-energy run and (b) for the high-energy run.  There are
approximately $2.44\cdot 10^6$ ($1.32\cdot 10^6$) $\pi^0$ events in the
low-energy and $572\cdot 10^3$ ($144\cdot 10^3$) $\pi^0$ events in the
high-energy data set. The numbers in parentheses give the number of
events with a proton seen by the Crystal-Barrel detector. This fraction
is larger for the high-energy run where the tagged energy
range started at higher energy, and more protons escaped through the
forward hole. For a fraction of this data, a more restrictive
trigger excluded events with only 2 detected particles.

The $\pi^0$~meson is observed above an almost negligible background at
a level of $10^{-3}$. The remaining background under the $\pi^0$ at
high energies was subtracted using side bins, which contained typically
a few events. Empty-target runs were used to determine additional
background. After all cuts had been applied, very few events survived
in empty-target runs. In LH$_2$ data, only $2\pm2$ \% were background
events not stemming from LH$_2$.

Cuts in the invariant mass were applied  in addition to the kinematic
fits. The criterion for a $\pi^0$ meson was a $2\gamma$ mass in the
($135\pm$35)\,MeV/$c^2$ interval.  At high photon energies, the cut was
widened to the 75 -- 175 MeV/$c^{2}$ region. For the
low-energy run, a second kinematic fit was applied constraining the
two-photon invariant mass to the $\pi^0$ mass (two-constraint fit).

\subsection{Monte-Carlo Simulations}

The detector was simulated by {\sc cb-geant}, a Monte Carlo program
based on the {\sc CERN} program package {\sc geant3}. The geometry of
the barrel calorimeter was implemented accurately. Only the type-11
crystals  (see fig. \ref{cb_calorimeter})
had to be approximated since {\sc geant3} did not support
their non-trapezoidal shape.

A number of small energy correction factors were applied to the
simulated energy deposits in the CsI crystals. Photons of larger energy
have a shower profile penetrating deeper into the crystals. The
efficiency of light collection is higher by several per cent at the
rear end of a crystal. In addition, the light collection efficiency is
different for different crystal shapes. The mean energy loss due to
shower leakage into support material was determined from the
simulations. These effects required empirical corrections of the
reconstructed photon energies.

\begin{figure*}
\hspace*{-0.5cm}\subfigure[\pir,1.4\,GeV]
{\epsfig{file=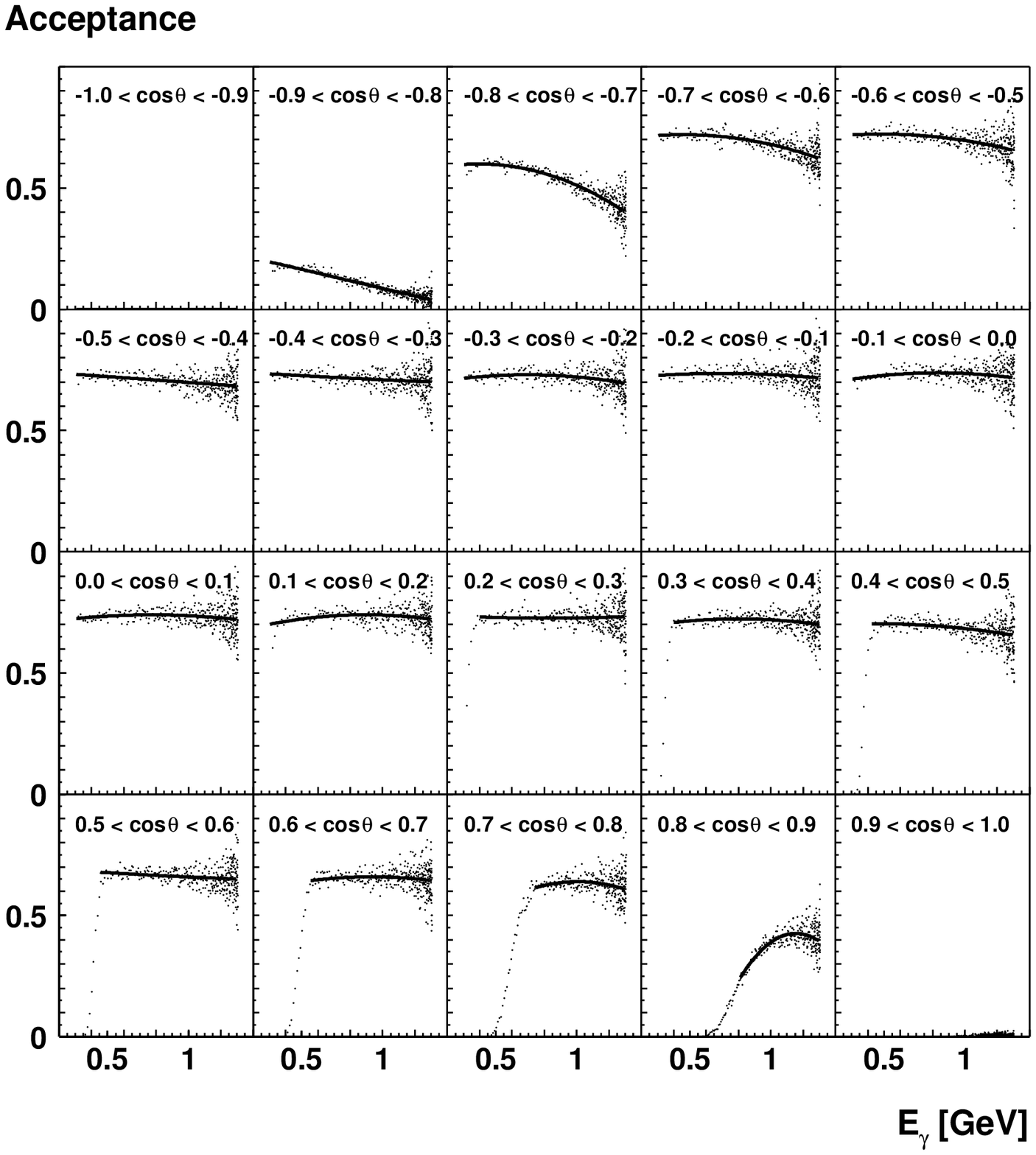,width=0.57\textwidth}}
\hspace*{-1.1cm}\subfigure[\pir, 3.2\,GeV]
{\epsfig{file=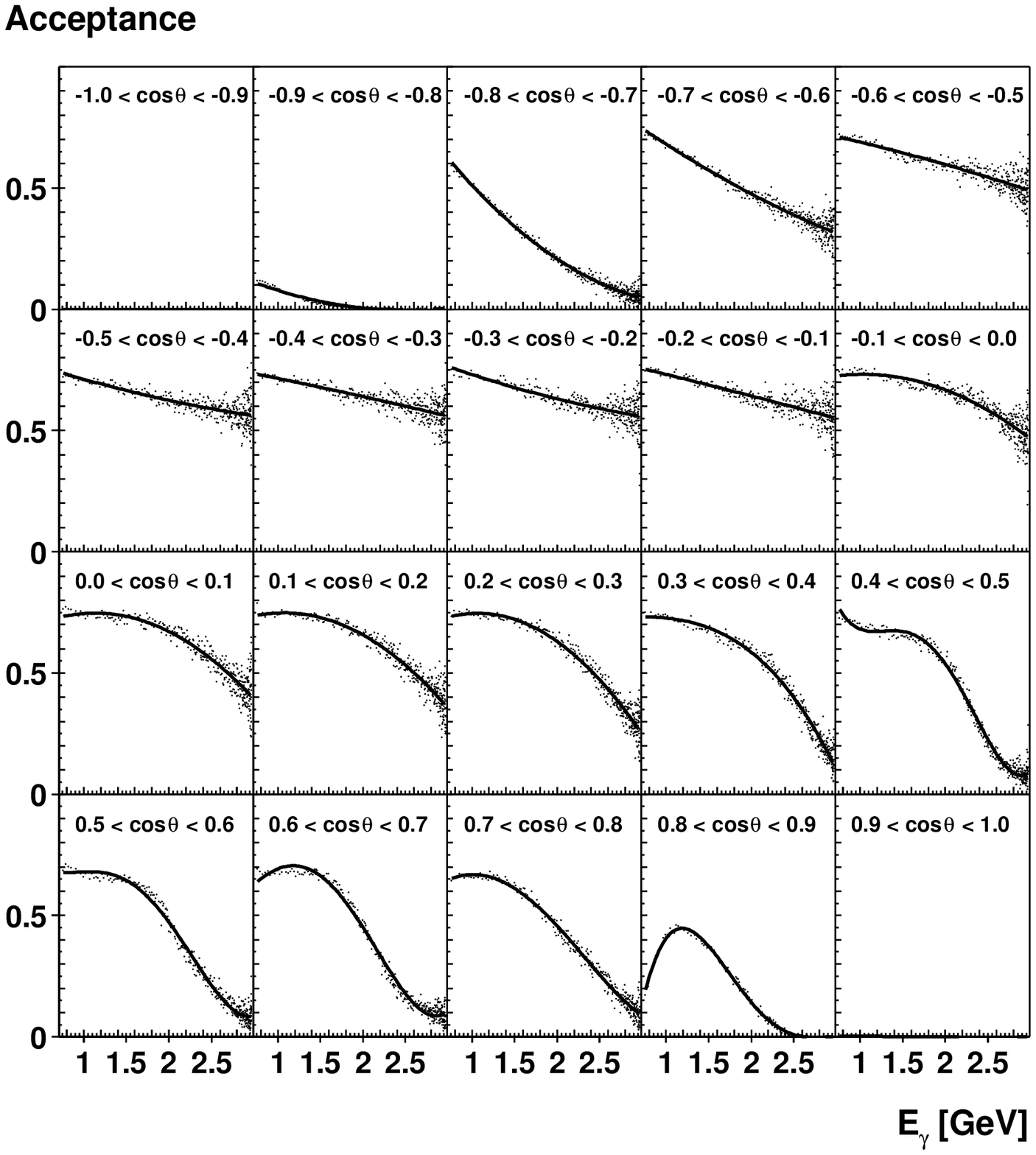,width=0.57\textwidth}}\\
\caption{Acceptance as a function of tagger wire number for the 20
$\cos\Theta_{cms}$ bins from $[-1.0,0.9],\ldots,[0.9,1.0]$. The
acceptances are plotted at the energy $E_{\gamma}$ of the
corresponding wire. The acceptance is calculated using
eq.~(\ref{eq:acc}). At large energies, the energy range covered by one
wire is small, and the acceptances show statistical fluctuations due to
limited Monte Carlo event numbers. } \label{FigureAcceptance}
\end{figure*}

The fiber detector was implemented as a homogeneous cylindrical
scintillation counter since simulating fibers with helix-shaped bending
is not supported by {\sc geant3}. Instead, the fiber number in each
layer was calculated from the known impact point taking into account
the real shape of the fibers. Single-fiber inefficiencies as derived
from the data were taken into account to allow simulations of trigger
efficiencies. The gaps between the cylinders were filled with a
carbon-based support structure. The plastic-foil wrapping was
simulated as described in (5.3.1).

The acceptance was determined from Monte-Carlo simulations as
ratio of reconstructed and generated events:
\begin{equation}
\label{eq:acc}
A_{\pi^0\rightarrow 2\gamma}=\frac{N_{\rm rec,MC}}{N_{\rm gen,MC}}.
\end{equation}
The acceptance is shown in Fig.~\ref{FigureAcceptance} for the 20
$\cos\Theta_{\rm cm}$ bins and for both data sets as a function of the
wire number. Instead of the wires number, the corresponding photon
energy is plotted on the horizontal axis. $A_{\pi^0\rightarrow
2\gamma}$ reaches a maximum of 60\% to 70\% depending on the energy
range. The acceptance is very small for $\cos \Theta_{\rm cm}>0.9$. The
corresponding low-energetic protons hardly reach the fiber detector
and thus the first-level trigger condition is not always met. For
$\cos\Theta_{\rm cm}<-0.9$ (for backward $\pi^0$), protons simply
escape through the forward hole not covered by the fiber detector. For
high-energy $\pi^0$ mesons, the opening angle between the two decay
photons is not large enough, and only a single PED is detected. Then
the event is lost in the trigger and the detection efficiency is
smaller at higher energies. To avoid biased results due to
imperfections of the apparatus and/or the simulations, we required the
detection efficiency to exceed a minimum of 5 \%. For this reason, we
rejected the last forward and backward data points.
\section{\label{SectionCrossSections}
Determination of Cross Sections}
 \subsection{Basic Definitions}
The unpolarized differential cross sections $\rm d\sigma/d\Omega$ can
be calculated from the number of data events identified in the respective
channel using
\begin{equation}
\frac{\rm d\sigma}{\rm d\Omega}=\frac{N_{{\rm \pi^0}\rightarrow
2\gamma}}{A_{{\rm \pi^0}\rightarrow
2\gamma}}\medspace\frac{1}{N_{\gamma}\rho_{\rm
t}}\medspace\frac{1}{\Delta\Omega}\medspace
\frac{\Gamma_{\rm total}}{\Gamma_{{\rm \pi^0}\rightarrow 2\gamma}},
\label{EquationDSigmaDOmega}
\end{equation}
where the quantities are:

\vspace{2mm}

\begin{tabular}{cl}
\hspace{-3mm}$N_{\pi^0\rightarrow 2\gamma}$
&:\quad  \text{Number of events in ($E_{\gamma},\cos\Theta_{\rm cm}$)
bin},\\
\hspace{-3mm}$A_{\pi^0\rightarrow 2\gamma}$&:\quad  \text{Acceptance in
($E_{\gamma},\cos\Theta_{cm}$)  bin},\\ \hspace{-3mm}$N_{\gamma}$ &:\quad
\text{Number of primary photons in $E$ bin},\\ \hspace{-3mm}$\rho_{\rm t}$ &:\quad
\text{Target area density},\\ \hspace{-3mm}$\Delta\Omega$ &:\quad \text{Solid-angle
interval } $\Delta\Omega=2\pi\Delta (\cos\Theta_{\rm cm})$,\\
\hspace{-3mm}$\frac{\Gamma_{\pi^0\rightarrow 2\gamma}}{\Gamma_{\rm
total}}$&:\quad  \text{decay branching ratio}.
\end{tabular}
\vspace{2mm}

The number of events in a $(E_{\gamma},\cos\Theta_{\rm cm})$ bin
comprises events with two or three PEDs in which the proton was
either undetected or detected in the Crystal Barrel. Summation
of both contributions reduces the necessity of reproducing exactly the
threshold behavior of low-energetic protons in the Crystal-Barrel
detector. For part of the 3.2-GeV data run, three or more hits in
the Crystal Barrel were already required by the hardware trigger
(FACE). The most forward data point has thus a smaller number of events
and a larger error.

The target area density is calculated from the density of liquid
hydrogen $\rho({\rm LH}_2)=0.0708\,$g/cm$^3$ and its molar mass $M_{\rm
mol}({\rm LH}_2)=2.01588\,$g/mol to be
\begin{align}
\rho_{\rm t} \ = \ \frac{2\rho({\rm LH}_2)N_{\rm A}L}{M_{\rm
mol}({\rm LH}_2)} \ = \ 2.231\cdot 10^{-7}/\mu{\rm b},
\end{align} where $N_{\rm A}=6.022\cdot 10^{23}\,$/mol is the
Avogadro's number, $l=52.75$\,mm is the length of the active target
cell, and the factor 2 accounts for the two atoms in ${\rm LH}_2$.

The $\pi^0$ was identified via its decay into $2\gamma$ which has a
relative branching ratio of 98.798 \%.

The solid-angle interval is
\begin{equation}
\Delta\Omega=2\pi\,\Delta(\cos\Theta_{\rm cm})\,,
\end{equation}
where $\Delta(\cos\Theta_{\rm cm})=0.1$ gives the bin width of the angular
distributions, subdividing $\cos\Theta_{\rm cm}$ into 20 bins.
Photon-energy bins of about 25\,MeV width were chosen for the 1.4\,GeV
data. The 3.2-GeV data are presented in bins of about 50\,MeV,
100\,MeV, and 200\,MeV in the intervals $E_{\gamma}\in[750,2300],
[2300,2600], [2600,3000]$, respectively.

\subsection{\label{Normalization}
Normalization}
At $E_{\rm e,~ELSA} = 1.4$\,GeV, the SAID-SM02 model description of
$\rm\gamma p\to p\pi^0$ was used for the absolute normalization of our
measured angular distributions. We minimized the function
\begin{equation}
\sum_{i=0}^{n-1}\frac{(x\cdot N_i-S_i)^2}{(\delta_{N_i})^2}
\end{equation}
for each energy channel of the tagging system, where $i$ denotes  a
bin in $\cos\Theta_{\rm cm}$, $N_i$ are the
acceptance-corrected number of data events, $\delta_{N_i}$ are the
errors in $N_i$, and $S_i$ are the values of the SAID model integrated
over the corresponding energy bins. Furthermore, $x$ stands for a
normalization factor including all normalization constants of
Eq.~(\ref{EquationDSigmaDOmega}), it is used as free fit parameter.
This method works well for the energy region between threshold and
$E_\gamma\approx 2$\,GeV, where the SAID model is reliable. The
extracted photon flux leads also in the channel $\rm \gamma p\to p\eta$
to cross sections which are consistent with previous data. We estimate
the error of the normalization to be $\pm 5$\%. The error given by the
statistical spread of all data on $\pi^0$ photoproduction used by SAID
is not taken into account.

The method of normalizing to a known cross section was not applicable
to our data at $E_{\rm e,~ELSA} = 3.2$\,GeV since our data extend the
currently available world database substantially. Instead, we relied
on the measured number of electrons in the tagger,  which is
proportional to the photon flux. A normalization factor was
determined from a fit to SAID-SM02 in the energy range up to 1.7\,GeV
and applied to the full energy range. Some systematic deviations in the
absolute cross section of the order of 10-15\,\% occur in the lowest
bins, for 800 - 950\,MeV.  The deviations are not caused by
uncertainties in detection efficiency. This follows from the perfect
agreement of the $2\gamma$ and $3\pi^0$ decay channels of the $\eta$ in
$\rm\gamma p\to p\eta$~\cite{Crede:2003ax}. Thus, we believe that
within a systematic error of 15\,\% we can trust this normalization,
even above incoming photon energies of 2\,GeV.

\subsection{Systematic Uncertainties}
\label{SectionSystematics}

Systematic errors were studied in Monte-Carlo simulations. The errors,
with exception of the normalization error, were added quadratically for
each point in the differential cross section. The total systematic
error is then added quadratically to the statistical error. The
systematic errors discussed in the following subsection are included in
the shown error bars. The
statistical errors are small, except for the most forward and most
backward points in the angular distributions.

The following effects contribute to the systematic
uncertainty of the measurements:

\subsubsection{Reconstruction Efficiency}
\vspace{-2mm}

The reconstruction of neutral mesons and the identification of final
states required a sequence of cuts including those on the results of
kinematic fitting. An overall uncertainty of $\pm 5.7$\% was assigned
to the reconstruction efficiency as determined in \cite{Amsler:1993kg}.
This error includes uncertainties due to split-offs.

\vspace{-4mm}

\subsubsection{Target Position}
\vspace{-2mm}

 The position of the target cell was determined by comparing results
from kinematic fitting (off-zero displacement of pull distributions) to
Monte-Carlo simulations. The target was found to be shifted upstream by
0.65\,cm, i.e. into the direction of the tagger with respect to the
center of the Crystal Barrel. The acceptances of the \pir~reaction were
re-determined for target shifts of $\pm 3$\,mm compared to the real
target position. The variations in the differential cross sections that
resulted from these changes depend on energy and angle are as large as
5\% for forward protons and $\pm1$\% on average.

\vspace{-4mm}

\subsubsection{Position of Beam Axis}
\vspace{-2mm}

The position of the photon beam measured by a beam profile
monitor showed slight variations in time and even within an
extraction cycle. Analyzing data from the inner detector, this shift
was found to be less than 3\,mm off axis at the target position. The
acceptances were re-calculated assuming a shifted vertex distribution
and the resulting cross section changes were included as systematic
uncertainty. The errors due to beam shifts are, depending on energy and
angle, 2\% or smaller.

\vspace{-4mm}

\subsubsection{Material between Target and Crystal Barrel}
\vspace{-2mm}

All material between target and sensitive detector components was
simulated carefully. However, black tape and foil was used to wrap
the inner detector to get it light-tight; its thickness was
not exactly known. Small changes in the thickness of this material had
large effects on the resulting detection efficiency for low energy
protons and on the trigger efficiency (which required at
least two layers of the scintillation fiber detector to have fired).
Assuming the worst case of 1\,mm of additional or missing material,
simulations were performed, new acceptances calculated and systematic
uncertainties in the cross sections determined. This error was
negligible. \vspace{-4mm}

\subsubsection{Solid Angle}
\vspace{-2mm}

Events belonging to one bin in $\cos\Theta_{\rm cm}$ may be reconstructed
in an adjacent bin. This {\it migration} effect was studied by
unfolding the resolution in $\cos\Theta_{\rm cm}$ by Monte Carlo
methods and found not to contribute significantly to the results.

\vspace{-3mm}

\subsubsection{Target Thickness}

The target thickness does not contribute to the
error since our cross sections are normalized to SAID.

\vspace{-4mm}

\subsubsection{Photon Flux}
\vspace{-2mm}

Our angular distributions are normalized to SAID, as outlined in
section \ref{Normalization}. We assign an error of $\pm 5$\% ($\pm
15$\%) to the normalization of the 1.4\,GeV (3.2\,GeV) data. These
errors are not included in Figs.
\ref{FigurePi0DCS14}-\ref{FigurePi0DCS}.

\section{Experimental Results}
\label{SectionResults}

Differential cross sections were calculated separately for the two
different ELSA energies of 1.4\,GeV and 3.2\,GeV. The data sets
published in \cite{Bartholomy:2004uz} contained the complete 1.4-GeV
data set covering the photon energy range from 0.8 to 1.3\,GeV and the
1.3 to 3.0\,GeV range from the 3.2-GeV data set.
\vspace{-2mm}
\subsection{Differential Cross Sections
\boldmath$\bf\rm
  d\sigma/d\Omega$\unboldmath\ for \pirbf~at an Electron-Beam
  Energy of 1.4 GeV}
\begin{figure}[pb]
\vspace{-5mm}
\begin{center}
\epsfig{file=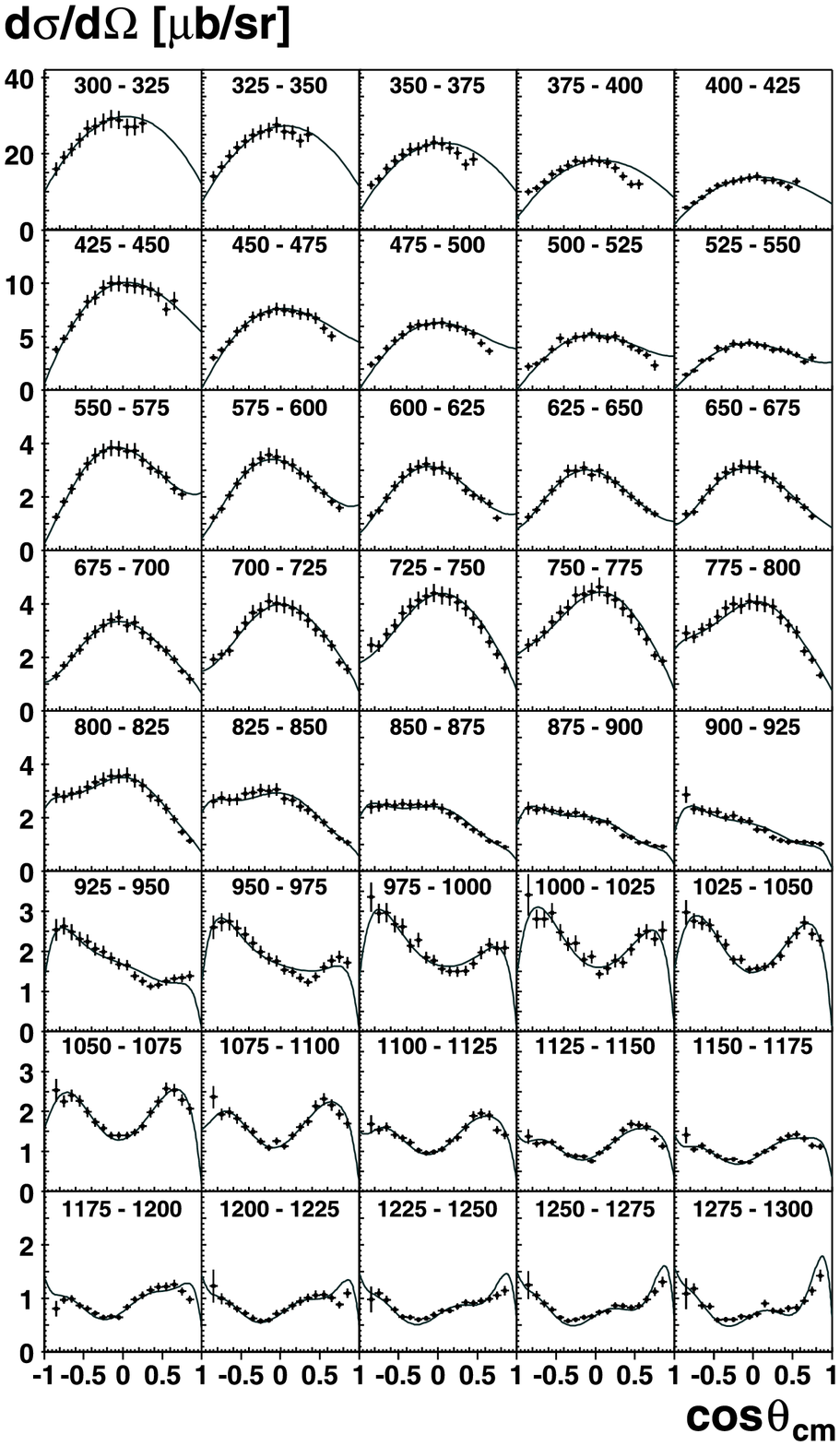,width=0.49\textwidth,height=0.59\textheight}
\end{center}
\caption{Differential cross sections for \pir~from the low-energy data.
Statistical and systematic errors are added quadratically.  The
photon flux was determined by normalizing the distributions to the
SAID-SM02 model.  The photon energy range (in MeV) is given in
each subfigure. The symbols are:
  {\tiny\ding{110}}: CB-ELSA, {\color{gray05}\bf---}: SAID-SM02.}
\label{FigurePi0DCS14}
\vspace{3mm}
\end{figure}

The differential cross sections are shown in Fig.~\ref{FigurePi0DCS14}
together with the SAID-SM02 model curve. Statistical and
systematic errors are added quadratically. In general,
the agreement is impressive. Note, however, that the data are
normalized to SAID in each energy bin. At the lowest energies, we
encounter some deviations from SAID. We cannot exclude systematic
effects beyond those listed in section \ref{SectionSystematics}. They
may be due to the rather small kinetic energies of the recoil protons
from the reaction \pir, just at the threshold to trigger the inner
detector. The high tagger rates in this energy region may also be
responsible for the problems.

\subsection{Differential Cross Sections \boldmath$\bf\rm
  d\boldsymbol\sigma/d\boldsymbol\Omega$\unboldmath ~for
\pirbf~at 3.2 GeV}

\begin{figure}[pb]
\begin{center}
\epsfig{file=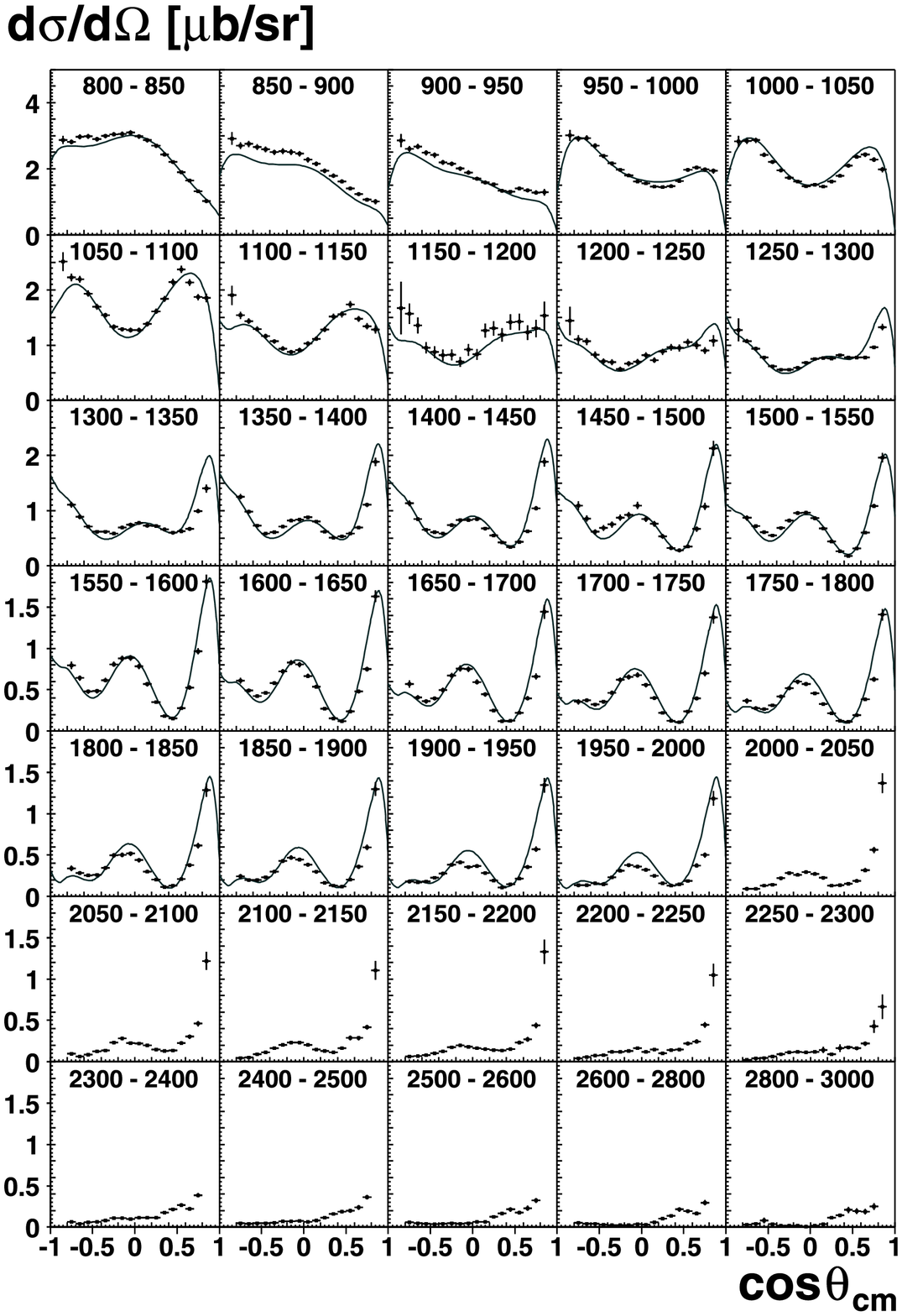,width=0.49\textwidth,height=0.52\textheight}
\end{center}
\caption{Differential cross sections for \pir~from the high-energy
data. Statistical and systematic errors are added quadratically. The
absolute normalization was determined by fitting the angular
distributions in the range from 800\,MeV to 1.7\,GeV to the SAID-SM02
model.  The photon energy range (in MeV) is given in each subfigure.
The SAID-SM02 model result is shown as solid line.}
\label{FigurePi0DCS32} \end{figure}

Fig.~\ref{FigurePi0DCS32} shows the angular distributions for the
reaction \pir~for the high-energy data set. SAID-SM02 results
are shown for comparison. There is good overall agreement
between data and SAID. At the smallest energies, the agreement is
somewhat worse. For this data, one normalization factor is used.

The two data sets are combined by using the 1.4-GeV data set up to the
maximum possible photon energy of 1.3\,GeV and using the 3.2\,GeV
data set to cover the range from 1.3\,GeV to 3\,GeV photon energy. We
restrained from calculating mean values for $0.8\leq E_{\gamma}\leq
1.3$\,GeV since the final errors are dominated by common systematic
errors. For this energy region, we used the 1.4-GeV data because
of the finer binning and since the tagger worked more reliably in the
high energy part where the intensity is lower. \vspace{-5mm}

\subsection{Differential Cross Sections
\boldmath$\rm d\sigma/d\Omega$\unboldmath\ for \pirbf, Combined Data
Set}

The cross sections from threshold to 3\,GeV are shown in
Fig.~\ref{FigurePi0DCS}. Data points are shown as open circles with
error bars. The solid line represents a fit described below. The MAID
results~\cite{Drechsel:1999} for photon energies up to 1\,GeV  are given
by dotted lines, results of the new SAID model SM05 for photon
energies up to 2\,GeV as dashed lines. The agreement between data and
the models is equally good for both data sets. The angular
distributions show rapid variations depending on both energy and
emission angle of the meson, reflecting the contributions from many
partial waves.

\epsfig{file=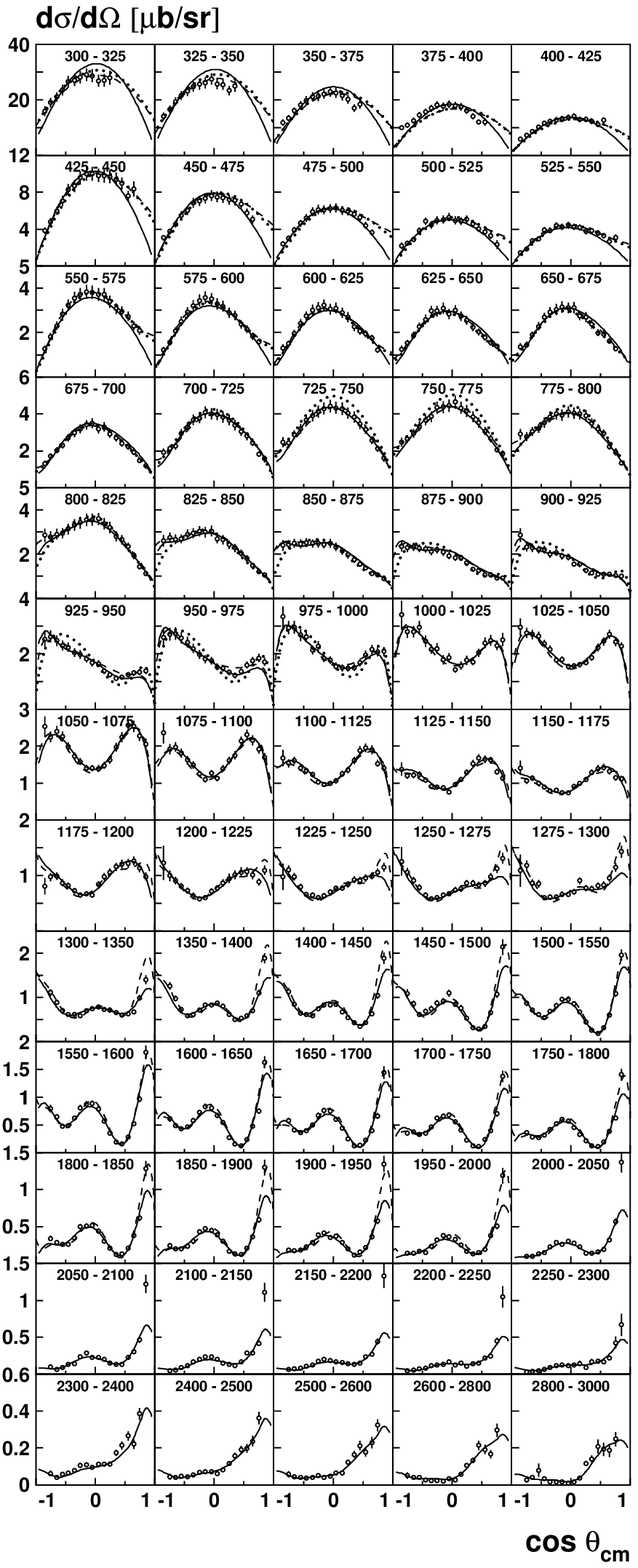,width=0.47\textwidth,height=1.02\textheight}
\begin{figure}[pt]
\caption{Previous page: Differential cross sections for \pir; combined
data set. Statistical and systematic errors are added quadratically.
The solid line represents a fit to
the partial waves described below, the dashed line is the SAID SM05,
the dotted line the MAID model. } \label{FigurePi0DCS}
\end{figure}

\subsection{Partial Wave Analysis} \label{SectionPWA}

The strong variations in the differential cross sections indicate strong
contributions of various resonances with different quantum numbers. To
determine quantum numbers and properties of contributing resonances,
the differential $\rm\gamma p \to p \pi^0$ cross sections were used in
a partial wave analysis. The analysis was based on an isobar model. In
the $s$-channel, $\rm N^*$ and $\rm \Delta^*$ contribute to the $\rm p
\pi^0$ final state. The background in this channel was described by
reggeized $t$-channel $\rho$\,($\omega$) exchange and by baryon
exchange in the $u$-channel. In addition $s$-channel Born-terms were
included in the fits. Details on the partial wave analysis can be found
in~\cite{Anisovich:2004zz,Anisovich:2005tf}. The CB-ELSA data on
$\rm\gamma p \to p \pi^0$ and $\rm\gamma p \to p\eta$
\cite{Bartholomy:2004uz,Crede:2003ax} were included as well as
additional data sets from other experiments: Mainz-TAPS data
\cite{Krusche:nv} on $\eta$ photoproduction, beam-asym\-metry
measurements of $\pi^0$ and $\eta$~\cite{Bartalini:2005wx,SAID1,SAID2},
and data on $\rm\gamma p\rightarrow n\pi^+$~\cite{GRAAL2}. The high precision
data from GRAAL~\cite{Bartalini:2005wx} do not cover the low mass region;
therefore we extracted further data from the compilation of the  SAID
database~\cite{SAID1}. Data on photoproduction of $\rm K^+\Lambda$,
$\rm K^+\Sigma^0$, and $\rm K^0 \Sigma^+$ from SAPHIR
\cite{Glander:2003jw,Lawall:2005np} and CLAS~\cite{McNabb:2003nf}, and beam asymmetry
data for $\rm K^+\Lambda$, $\rm K^+\Sigma^0$ from LEPS
\cite{Zegers:2003ux} were also included in the analysis. This partial
wave analysis is much better constrained than an analysis using the
$\rm\gamma p \to p \pi^0$ data only. The main results on baryon
resonances coupling to $\rm p \pi^0$ and $\rm p\eta$ are discussed
in~\cite{Anisovich:2005tf}, those to $\rm K^+\Lambda$ and $\rm
K^+\Sigma^0$ are documented in~\cite{Sarantsev:2005tg}.

To describe the different data sets, 14\, N$^*$ resonances coupling to
N$\pi$, N$\eta$, $\rm K\Lambda$, and  $\rm K\Sigma$ and 7 $\rm \Delta^*$
resonances coupling to N$\pi$ and $\rm K\Sigma$ were needed. Not all
included resonances contribute to the $\rm p\pi^0$ final state.
Most resonances were described by relativistic Breit-Wigner
amplitudes. For the two S$_{11}$ resonances at 1535 and 1650\,MeV, a
four-channel $K$-matrix ($\rm N\pi$, $\rm N\eta$, $\rm K\Lambda$,
$\rm K\Sigma$) was used.

The total cross section and the main results of the PWA are briefly
discussed in the next section.

\subsection{Total Cross Section and Results of the Partial Wave Analysis}
\begin{figure}[pt]
\epsfig{file=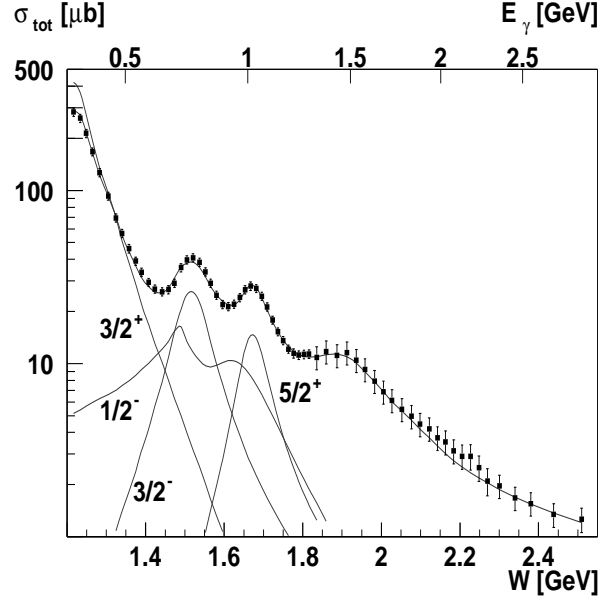,width=0.49\textwidth,height=0.45\textwidth,clip=}
\caption{\label{figure_pi0tcs} Total cross section (logarithmic
scale) for the reac\-tion $\rm\gamma p\rightarrow p\pi^0$ obtained
by integration of angular distributions of the CB-ELSA data and
extrapolation into forward and backward regions using our PWA
result. The solid line represents the result of the PWA. Four
individual contributions to the cross section are also shown.}
\end{figure}

From the differential cross sections, the total cross section was
determined by integration. The integration was performed by summing over
the differential cross sections of Fig. \ref{FigurePi0DCS} and using
extrapolated values from the fit for bins with no data. In the total
cross section, shown in Fig.~\ref{figure_pi0tcs}, clear peaks are
observed for the first, second, and third resonance region. The fourth
resonance region exhibits a broad enhancement at W about
1900\,MeV. The decomposition of the peaks into partial waves and their
physical significance will be discussed below.

The first resonance region is the dominant structure of
Fig.~\ref{figure_pi0tcs}. It is due to excitation of the
$\rm\Delta(1232)P_ {33}$. There is strong destructive interference
between $\rm\Delta(1232)$ $\rm P_ {33}$, the $\rm P_{33}$ nonresonant
amplitude, and $u$-channel exchange. The $\rm
N(1440)P_ {11}$ Roper resonance provides only a small contribution of
about 1--3\% compared to the $\rm \Delta(1232){\rm P}_{33}$.
In the second resonance region, the $\rm N(1520)D_{13}$ and
$\rm N(1535) S_{11}$ resonances yield contributions which
are shown as thin lines in Fig. \ref{figure_pi0tcs}.

 The third bump in the total cross section is due to three
major contributions: the $\rm\Delta(1700)D_{33}$ resonance provides the
largest fraction ($\sim\! 35$\%) of the peak, followed by $\rm
N(1680)F_{15}$ ($\sim\! 25$\%) and $\rm N(1650)S_{11}$ ($\sim\! 20$\%).
In addition the $\rm\Delta(1620)S_{31}$ ($\sim\! 7$\%) and $\rm
N(1720)P_{13}$ ($\sim\! 6$\%) resonances are required.  In the fourth
resonance region, the $\rm\Delta(1950)F_{37}$ contributes $\sim\! 41$\%
to the enhancement and $\rm\Delta(1920)P_{33}$ is identified with
$\sim\! 35$\%. Additionally, the fit requires the presence of
$\rm\Delta(1905)F_{35}$ and $\rm\Delta(1940)D_{33}$.
The high-energy region is dominated by $\rho$($\omega$) exchange in the
$t$-channel as can be seen by the forward peaking in the
differential cross sections.

\subsection{Differential Cross Sections \boldmath$\bf
d\sigma/dt$\unboldmath} \label{SectionTDistributions} The partial
wave analysis assigns a fraction of the total cross section to
$t$-channel exchange which increases with energy. This was already
expected from the strong rise of the differential cross sections
towards $\cos(\Theta_{cm})\sim 1$ in the higher energy bins.

\begin{figure}[pt]
\begin{center}
\epsfig{file=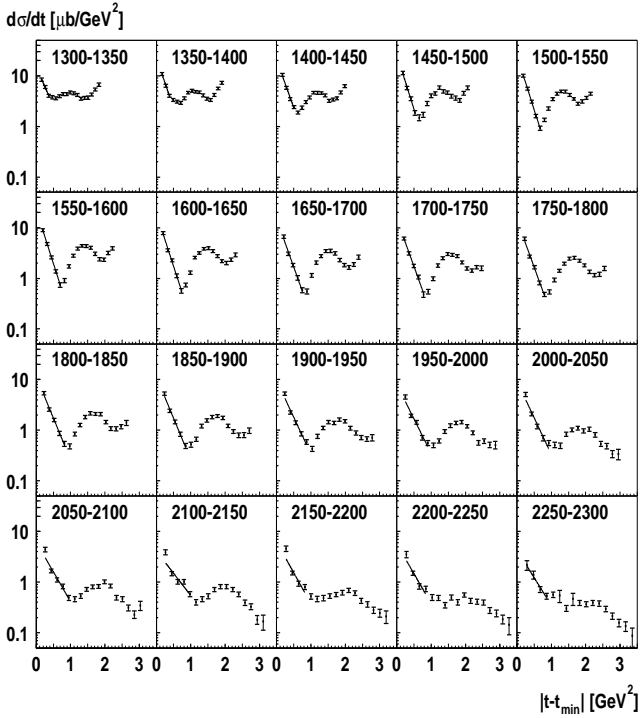,width=0.47\textwidth,height=0.53\textwidth}
\end{center}
\caption{Differential cross sections $\rm d\sigma$/${\rm d}t$ for \pir.
In the range of small values of $|t-t_{\rm{min}}|$ the data were fitted
with an exponential function $\rm{e}^{a+b|t-t_{\rm{min}}|}$ distribution.}
\label{FigureDSigmaDT}
\end{figure}

Fig. \ref{FigureDSigmaDT} shows differential cross sections as functions
of the squared four-momen\-tum transfer between the initial photon and
the $\pi^0$ in the final state. Even though these plots do not provide
new information compared to $\rm d\sigma/d\Omega$, they are shown here
to emphasize the exponential fall-off of the cross section at low $t$.
The differential cross sections are plotted against $|t-t_{\rm min}|$,
the difference between the actual momentum transfer and the minimal
momentum transfer imposed by kinematics. By definition, the
four-momentum transfer is always negative and can range between $t_{\rm
min}$ and $t_{\rm max}$, which for $\gamma$p experiments is given by
\begin{align}
t_{\rm{min/max}}& = \\
 \left[\frac{m_{\rm{\pi^0}}^2}{2\sqrt{s}}\right]^2
&-\left[\frac{s-m_{\rm{p}}^2}{2\sqrt{s}}\mp\sqrt{\frac{(s+m_{\rm{\pi^0}}^2
-m_{\rm{p}}^2)^2}{4\,s}-m_{\rm{\pi^0}}^2}\,\,\right]^2.\nonumber
\label{EquationTMinTMaxSimpler}
\end{align}
Note that $s=m_{\rm p}^2+2m_{\rm p}E_{\gamma}$ and that
$t_{\rm min}$ corresponds to forward, $t_{\rm{max}}$ to backward production
of the $\pi^0$.

The squared momentum transfer is related to the emission angle of the pion
in the center-of-mass system by
\begin{align}
t=t_{\rm{min}}-\nonumber
&\frac{s-m_{\rm{p}}^2}{s}
\sqrt{(s-(m_{\pi^0}+m_{\rm{p}})^2)}\nonumber\\
&\sqrt{(s-(m_{\pi^0}-m_{\rm{p}})^2)}
\sin^2\frac{\Theta_{\rm{cms}}}{2}\,.
\end{align}

\noindent
Differential cross sections d$\sigma$/d$\Omega$ are related to
d$\sigma$/d$t$ by
\begin{equation}
\frac{\rm d\sigma}{\rm d\Omega}=\frac{\rm d\sigma}{{\rm d}t}\frac{{\rm
    d}t}{{\rm d}\Omega}\,.
\label{eqn23}
\end{equation}

Using (\ref{eqn23}), the relation of the two different differential
cross sections is given by:
\begin{equation} \frac{\rm d\sigma}{{\rm
d}t}= \frac{4\pi~s~{\rm d\sigma /d\Omega}}
{(s-m_{\rm{p}}^2)\sqrt{(s-(m_{p}+m_{\rm{\pi^0}})^2)
(s-(m_{p}-m_{\rm{\pi^0}})^2)}}.
\label{EquationDSigmaDT}
\end{equation}

For small four-momentum transfers, an exponential function is fitted to
the distributions shown in Fig.~\ref{FigureDSigmaDT}. The cross
sections fall off according to
\begin{equation}
\label{expo}
\frac{{\rm d}\sigma}{{\rm d}t}\propto\exp(a+b\cdot|t-t_{\rm{min}}|),
\end{equation}
where the slope parameter $b$ has a negative value.

\begin{figure}[ph]
\vspace{-5mm}
\begin{center}
\epsfig{file=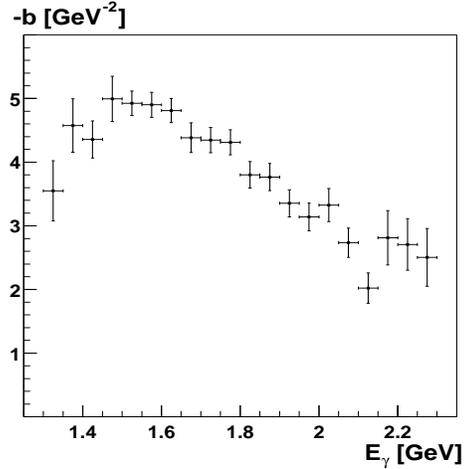,width=0.40\textwidth,height=0.35\textwidth}
\end{center}
\caption{Slope parameter for \pir from a fit using
$\rm{e}^{a+b|t-t_{\rm{min}}|}$ to the differential cross sections
$\rm d\sigma$/${\rm d}t$.}
\label{FigureSlopeParameter}
\end{figure}

For large $t$, the cross sections do not exhibit the behavior that
would be expected if we had only $t$-channel exchange of mesons.  There
are structures due to other phenomena, in particular due to formation
of $s$-channel resonances. The range, in which the distributions can be
fitted by an exponential is not well defined. Nevertheless, we show in
Fig.~\ref{FigureSlopeParameter} the slope parameter $-b$ of Eq.
\ref{expo} from the fits as a function of incident photon energy
$E_{\gamma}$. With increasing energy, the slope rises to a
maximum value at $E_{\gamma}\sim$1.5--1.6\,GeV and then decreases
again. At large energies, above 2.2\,GeV, the determination of the
slope parameter becomes somewhat arbitrary, since the most forward data
points at $\cos\Theta_{\rm cm}=0.85$ are missing. For low energies, the
slope parameter cannot be determined from the data and can even adopt
positive values. Obviously, resonance production is dominant at
these energies. The results should hence be interpreted with care.
In particular, the turnover at $E_{\gamma}\leq 1.5$\,GeV is deduced
from very few data points.

\section{Summary}
\label{SectionSummary}
We have reported a measurement of unpolarized differential cross
sections of the reaction \pir~ in the photon energy range from 0.3\,GeV
to 3.0\,GeV, thus completely covering the baryon resonance region.
Above 0.8\,GeV, the data supersede previous experiments both in solid
angle coverage and in statistics.  In the mass range above the first
resonance region, strong variations of the differential cross sections
as functions of photon energy and decay angle in the rest frame of the
intermediate state suggest that many partial waves contribute to
$\pi^0$ photoproduction off protons. This is confirmed by a partial
wave analysis which, beyond non-resonant contributions, identifies a
series of resonances contributing to the $\rm\gamma p\to p\pi^0$
reaction channel. At large photon energies, the angular distributions
show a forward rise in $\pi^0$ direction signaling dominance of
$t$-channel exchange contributions at high energies. The data provide
valuable input to coupled-channel isobar-model descriptions of baryons
and their excitations. A partial wave analysis reveales a rich spectrum
of resonant and non-resonant contributions.\\[0.1cm]

\subsection*{Acknowledgmentss}
We thank the technical staff at ELSA and at all the participating institutions
for their important contributions to the success of the experiment. We
acknowledge financial support from the Deutsche Forschungsgemeinschaft (DFG)
within the SFB/TR16. The collaboration with St. Petersburg received funds
from DFG and the Russian Foundation for Basic Research. B.~Krusche
acknowledges support from Schweizerischer Nationalfond.

U.~Thoma thanks
for an Emmy-Noether grant from the DFG.  A.\ V.\ Anisovich and A.\ V.\
Sarantsev acknowledge support from the Alexander von Humboldt
Foundation. This work comprises part of the PhD theses of O.~Bartholomy
and H. van Pee. \phantom{Dummy}


\end{document}